\documentclass[twocolumn]{aastex63}
\usepackage[colorinlistoftodos]{todonotes}
\usepackage{amsmath}
\newcommand{\be}{\begin{eqnarray}}
\newcommand{\ee}{\end{eqnarray}}
\usepackage{CJKutf8}
\usepackage{xcolor}
\begin{document}
	
\title{ WOCS 5379: Detailed Analysis of the Evolution of a Post-Mass-Transfer Blue Straggler  }
\begin{CJK*}{UTF8}{gbsn}
\author[0000-0001-9037-6180]{M. Sun(孙萌)}
\affiliation{Department of Astronomy, University of Wisconsin - Madison, 475 N Charter St, Madison, WI 53706, USA}
\author[0000-0002-7130-2757]{Robert D. Mathieu}
\affiliation{Department of Astronomy, University of Wisconsin - Madison, 475 N Charter St, Madison, WI 53706, USA}
\author[0000-0002-3944-8406]{Emily M. Leiner}
\affiliation{Center for Interdisciplinary Exploration and Research in Astrophysics (CIERA) and Department of Physics and Astronomy, Northwestern University, 2145 Sheridan Road, Evanston, IL 60208, USA}
\affiliation{NSF Astronomy and Astrophysics Postdoctoral Fellow}
\author[0000-0002-2522-8605]{R. H. D. Townsend}
\affiliation{Department of Astronomy, University of Wisconsin - Madison, 475 N Charter St, Madison, WI 53706, USA}

\begin{abstract}
	
The blue-straggler binary WOCS 5379 is a member of the old (6-7 Gyr) open cluster NGC 188. WOCS 5379 comprises a blue straggler star with a white dwarf companion in a 120-day eccentric orbit. Combined with the orbital period, this helium white dwarf is evidence of previous mass transfer by a red giant. Detailed models of the system evolution from a progenitor main-sequence binary, including mass transfer, are made using the Modules for Experiments in Stellar Astrophysics (MESA). Both of the progenitor stars are evolved in the simulation. WOCS 5379 is well reproduced with a primary star of initial mass 1.19 $M_{\odot}$, whose core becomes the white dwarf. The secondary star initially is 1.01 $M_{\odot}$. 300 Myr ago, the secondary finished receiving mass from the donor, having moved beyond the NGC 188 turnoff as a 1.20 $M_{\odot}$ blue straggler. The successful model has a mass transfer efficiency of 22\%. This non-conservative mass transfer is key to expanding the orbit fast enough to permit stable mass transfer. Even so, the mass transfer begins with a short unstable phase, during which half of the accreted mass is transferred. With increasing mass, the secondary evolves from a radiative core to a convective core. The final blue straggler interior is remarkably similar to a 2.1 Gyr-old 1.21 $M_{\odot}$ main-sequence star at the same location in the HR diagram. The white dwarf effective temperature is also reproduced, but the modeled white dwarf mass of 0.33 $M_{\odot}$ is smaller than the measured mass of 0.42 $M_{\odot}$.
	
\end{abstract}


\section{Introduction}
\label{sec:intro}

Blue straggler stars (BSSs) in clusters are stars that appear to be younger than the cluster age. Traditionally identified as brighter and to the blue of the main-sequence (MS) turnoff, BSSs can also be fainter than the turnoff but blueward of the MS. Similar stars also have been found within the MS \citep{2019ApJ...881...47L}. BSSs are thought to be MS stars that have gained mass, forming within binary stars or hierarchical multiple-star systems. The formation processes may be internal to the binary, via merger or mass transfer (MT) \citep{1964MNRAS.128..147M,1971ARA&A...9..183P}. Another formation channel, primarily in clusters, is formation through collisions within binary dynamical encounters \citep{1976ApL....17...87H}. Given the prevalence of binary stars, BSSs trace frequent alternative stellar evolution pathways to single-star evolution. 

Previous theoretical work discusses all three formation channels. \citet{2006ApJ...646.1160A} and \citet{2009Natur.457..288K} predict formation rates of single BSSs from close binary mergers. BSS merger formation can also happen in hierarchical triple-star systems via Kozai and tidal mechanisms merging inner binaries, producing BSSs in long-orbital-period systems, as suggested by \citet{2009ApJ...697.1048P} and further investigated by \citet{2011MNRAS.410.2370L} and \citet{2019MNRAS.488..728F}. 
For the MT scenario, \citet{1996ApJ...470..521L} and \citet{2004MNRAS.355.1182C, 2008MNRAS.387.1416C} analyze the feasibility of MT. They find that MT during the donor star's MS phase (Case A) yields very short-period binaries, rare among BSSs. Comparing BSSs from extremely close binaries with BSSs in the old open cluster M67, \citealt{2006A&A...455..247T} found similar results. MT during the red giant branch (RGB) phase yields BSSs in orbits with periods of hundreds of days and MT during the asymptotic giant branch phase (AGB) yields very-long-period binary BSSs, both of which are commonly found \citep{2009Natur.462.1032M}.

Recently ``blue lurkers'' have been found within the MS of the old open cluster M67. They were identified as fast rotators in wide binaries. Blue lurkers also may have merger, collision or MT origins, since all three mechanisms can spin up the rotation of the stellar product. The blue lurkers are thought to be less massive than classical BSSs \citep{2019ApJ...881...47L}. 

NGC 188 is a well-observed, old (6.2 Gyr) open cluster that is an excellent laboratory to test and further develop models for these mechanisms. Extensive precise radial-velocity time-series measurements of solar-like cluster members have been done by \citet[and references therein]{2008AJ....135.2264G}, providing an ideal sample for binary-evolution study. A large sample of orbit solutions are provided in \citet{2009AJ....137.3743G}. By analyzing the companion masses, the orbital periods, and the eccentricity distribution of 15 binary BSSs in NGC 188, \citet{2011Natur.478..356G,2012AJ....144...54G} conclude that the data favor MT as the most frequent formation pathway, and rule out the collision channel for the long-period binary BSSs. This conclusion was supported by the discovery of WD companions to 7 of the NGC 188 BSSs by  \citet{2014ApJ...783L...8G,2015ApJ...814..163G}, who conclude that approximately 2/3 of the NGC 188 BSSs formed through MT. More recently, \cite{2019ApJ...885...45G} measured the masses of two of these WDs via \textit{Hubble Space Telescope} COS spectra.

The goal of this paper is to provide the first detailed study of the formation of an NGC 188 BSS via the MT channel. The binary of choice is WOCS 5379, a BSS binary with a 0.4 $M_{\odot}$ helium WD companion in a 120-day orbit. The orbital period and the WD mass suggest that WOCS 5379 is an example of MT formation from a red giant \citep{2019ApJ...885...45G}. 

Although the origins of BSSs have been discussed for several decades, complete and detailed binary evolution models via the MT mechanism are rare. The only near-successful evolutionary model of a BSS-WD binary is that for V106 in NGC 6791 \citep{2018MNRAS.481.5062B}. This simulation of early Case B MT explains the age, very short orbital period (1.4 days), and most of the measurements of the post-MT white dwarf secondary star, but was not able to reproduce the BSS.

The recently available binary-evolution module within Modules for  Experiments  in  Stellar  Astrophysics (MESA; \citealt{2015ApJS..220...15P}) expands the possibilities for detailed MT studies. The organization of this paper is as follows. Section \ref{obs_fact} summarizes the observations of WOCS 5379. Section \ref{best_model} presents the MESA binary evolution model best reproducing WOCS 5379. Section \ref{bi_evol} discusses general physical issues driving this best-fit model. Section \ref{sec: stg_evol} presents in detail the binary evolution of the best-fit model.  Section \ref{discussion} explores the effects of secondary mass, MT efficiency, the location of the mass loss from the system , and orbital eccentricity on the binary evolution.
Section \ref{age_BS} presents a framework for discussion of BSS ages. Finally, the conclusions are given in Section \ref{conclusion}.

\section{Observed Properties of the WOCS 5379 System}
\label{obs_fact}

The binary nature of WOCS 5379 was discovered by \citet{2008AJ....135.2264G}, with a spectroscopic orbital solution presented by \citet{2009AJ....137.3743G}. As noted above, the 120.21-day period, combined with the WD detection, suggests a RGB MT origin. While the orbital eccentricity of 0.24 is no longer a surprise for post-MT systems (e.g., \citealt{2011Natur.478..356G}), a full explanation for the survival or generation of such eccentricity during the MT process is needed (see Section \ref{discussion}).

\citet{2015ApJ...814..163G} found WOCS 5379 to have the second hottest WD companion among the NGC 188 BSSs, with a photometric temperature of $17600\pm 500$ K and an inferred cooling age of only 77 Myr. WOCS 5379 sits very near the ZAMS to the blue of the evolved MS. It is the second-least luminous of the 21 known BSS in NGC 188. The BSS effective temperature is $6400 \pm 120~ {\rm K}$.

More recently, \citet{2019ApJ...885...45G} undertook a detailed analysis of a far-ultraviolet COS spectrum (central wavelength 1105 $\AA$) of WOCS 5379. Adopting a Gaia-based distance to NGC 188, they find the WD companion to have an effective temperature $T_{\rm eff}$ = 15,500 K and a surface gravity $\log\,g$ = 7.50 (cm s$^{-2}$). They find a mass for the secondary of $0.42\pm 0.02 M_{\odot}$ (where the uncertainty is a precision) and a cooling age of 250 Myr.

Finally, the presence of WOCS 5379 in a well-studied open cluster is of great benefit. \citet{2019ApJ...885...45G} review current distance estimates to NGC 188; following them we adopt a Gaia-based distance of 1845 $\pm$ 107 pc. The uncertainty encompasses other distance estimates. We also adopt $E(B-V)=0.09$ \citep{1999AJ....118.2894S} and solar metallicity \citep{1990AJ....100..710H,2011AJ....142...59J}. Age determinations for NGC 188 vary from 6.2 $\pm$ 0.2 Gyr to 7 $\pm$ 0.5 Gyr \citep{2009AJ....137.5086M,1999AJ....118.2894S}, while the Bayesian analysis of
\citet{2015AJ....149...94H} shows systematic uncertainties in NGC 188 age determinations of 0.7 Gyr. As a reference point, with MESA modeling we find the turnoff mass of NGC 188 to be 1.1 $M_{\odot}$.

Table \ref{observation vs model} lists these known quantities of the system and of the two stars.

\section{Numerical Simulation of WOCS 5379}
\label{best_model}
	
We undertake an evolutionary simulation for WOCS 5379 from an initial pair of MS stars to a BSS-WD pair, using the one-dimensional MESA code (Version 12115; \citealt{2011ApJS..192....3P,2013ApJS..208....4P,2015ApJS..220...15P,2018ApJS..234...34P,2019ApJS..243...10P}). The initially more massive primary star is the progenitor of the white dwarf. The initially less massive secondary star is the progenitor of the BSS, after accreting mass from the evolving primary star. Specifically, we compute simultaneously the evolution of both stars and the binary orbit in the presence of Roche Lobe overflow.
	
We use standard settings for MESA stellar evolution and binary evolution; our inlist is shared at xxxx.
Following \citet{2009AJ....137.3743G}, we adopt a metallicity of $Z=0.019$ in the simulation. For both stars, we set the outer boundary condition on the temperature and pressure via interpolation in pre-computed photosphere tables.
We do not include any stellar winds, as these stars have very little mass loss from the MS to the early RGB.

Previous study indicates that  non-conservative MT (where a fraction of the material leaving the donor star escapes from the system) can determine whether MT is stable or unstable \citep{2012ApJ...744...12W}. Our simulation allows non-conservative MT, with mass escaping via a fast wind from the vicinity of the accretor (see also Section \ref{MT_alpha_mechanism}). The MT efficiency, which is the percentage of the mass leaving the donor star that is accreted, is fixed in order to make the experiment feasible (see Section \ref{conclusion}). 
 
The material escaping from the system reduces the system angular momentum. However, the angular momentum per unit mass of the system increases during the mass loss, resulting in an increase in the system separation. 

Finally, as the accretor receives mass from the donor star, the simulation permits it to respond, for example by restructuring its radiative and convective zones.

This section first describes how we find the best-fit model for WOCS 5379 -- that is one that matches most of the observed parameters within the measurement uncertainties. (We note that the ``uncertainty" in the BSS luminosity is actually the range of brightness variation in the V band \citep{2003AJ....126..276K}.) Hereinafter we call this the Best-Fit (BF) Model. The initial parameters and why we start the search in a specific range are briefly discussed. Then the main features of the BF model and the comparison between the model and data are discussed in the last subsection.

\subsection{Finding the Best-Fit Model for WOCS 5379}

Preliminary MESA binary evolution trials with Z = 0.019 showed that a 1.2 $M_\odot$ donor star ends MT at $\approx$ 6.2 Gyr. \citet{2019ApJ...885...45G} find a cooling age of the WOCS 5379 WD of 250 Myr. Given an age for NGC 188 between 6.2 Myr and 7.0 Myr, we explore primary-star initial masses near 1.2 $M_\odot$.

We began our search of parameter space with a primary star mass of 1.2 $M_\odot$. The parameters searched were initial orbital period, initial secondary mass and MT efficiency. Since MT approaches instability with increasing mass ratio, our search started with a secondary mass slightly smaller than the primary mass. For each secondary star mass, MT efficiencies (ranging from 5\% to 100\%) and initial orbital periods (ranging from 10 to 50 days) were searched for models that came near to an orbital period of 120 days. The search continued with lower secondary masses until no stable MT occurred within this search space. The smallest such secondary mass was 0.9$M_{\odot}$. An upper limit of 40\% on the MT efficiency resulted from either MT instability or overly massive and luminous BSS products, depending on location in parameter space.
Finally, we made small adjustments in the parameters to best converge on the BSS observational data, and in system age to match the WD effective temperature.

No models were able to reproduce the WD $\log g$ measurement, and so we treat WD $\log g$ and mass as outcomes, not as independent parameters for goodness of fit.

Table \ref{initial model} displays the initial parameters of the BF model. The progenitor binary has an initial period of only 12.7 days. The mass transfer efficiency of 22\% is crucial in forming the observed 120-day period of the WOCS 5379 system. In this paper, all the physical quantities of the initially more massive star, which is the donor star and the WD progenitor, are assigned the subscript ``1''. The physical quantities of the initially less massive star that becomes the BSS have a subscript ``2''.

\subsection{The Best-Fit Model}

Table \ref{observation vs model} compares the Best-Fit Model against the observations. The model agrees well with most of the measured properties. The notable exception is the model WD $\log g$, which is higher than measured, yielding a less massive WD compared with the measured WD mass. No variations in model parameter space was able both to resolve this difference and to maintain acceptable fidelity to the other observed properties of WOCS 5379.

\begin{table*}[]
\caption{Properties of WOCS 5379 and the Best-Fit model}
\begin{center}
\begin{tabular}{c c c}
\hline\hline 
Property (Unit)	&	Observation& Best-Fit Model	\\
\hline
BSS $T_{\rm eff}$ (K) & 6400$\pm$120 & 6350\\
BSS Luminosity ($L_{\odot}$) & 	 2.5$\pm$0.5 &  2.6	\\
BSS Mass ($M_{\odot}$) & 	 -  &  1.20	\\
WD $T_{\rm eff}$ (K)& $15500\substack{+170 \\ -150}$& 15455 \\
WD $\log\,(g/(\rm cm\,s^{-1})$) &	$7.50\substack{+0.06 \\ -0.05}$ & 7.22	\\
WD mass ($M_{\odot}$) & 0.42 $\pm$ 0.02 & 0.33 \\
$P_{\rm orb}$ (day) & 120.21 $\pm$ 0.04 & 119.94 \\
Age (Gyr) & 6.2 - 7.0 & 6.77\\
Metallicity & 0.019 & 0.019 \\
Eccentricity & 0.24 $\pm$ 0.03 & 0  \\
\hline
\label{observation vs model}
\end{tabular}
\end{center}
\end{table*}

\begin{table*}[]
\caption{Initial settings for the best-fit model}
\begin{center}
\begin{tabular}{c c}
\hline\hline 
Initial conditions (Unit)	&	value	\\
\hline
$M_{\rm 1,i}$ ($M_{\odot}$) & 1.19\\
$M_{\rm 2,i}$ ($M_{\odot}$) & 1.01\\
$P_{\rm orb,i}$ (days) & 12.7\\
Mass Transfer Effeciency & 22\% \\
\hline
\label{initial model}
\end{tabular}
\end{center}
\end{table*}

\begin{figure}[tp]
\includegraphics[width=0.5\textwidth]{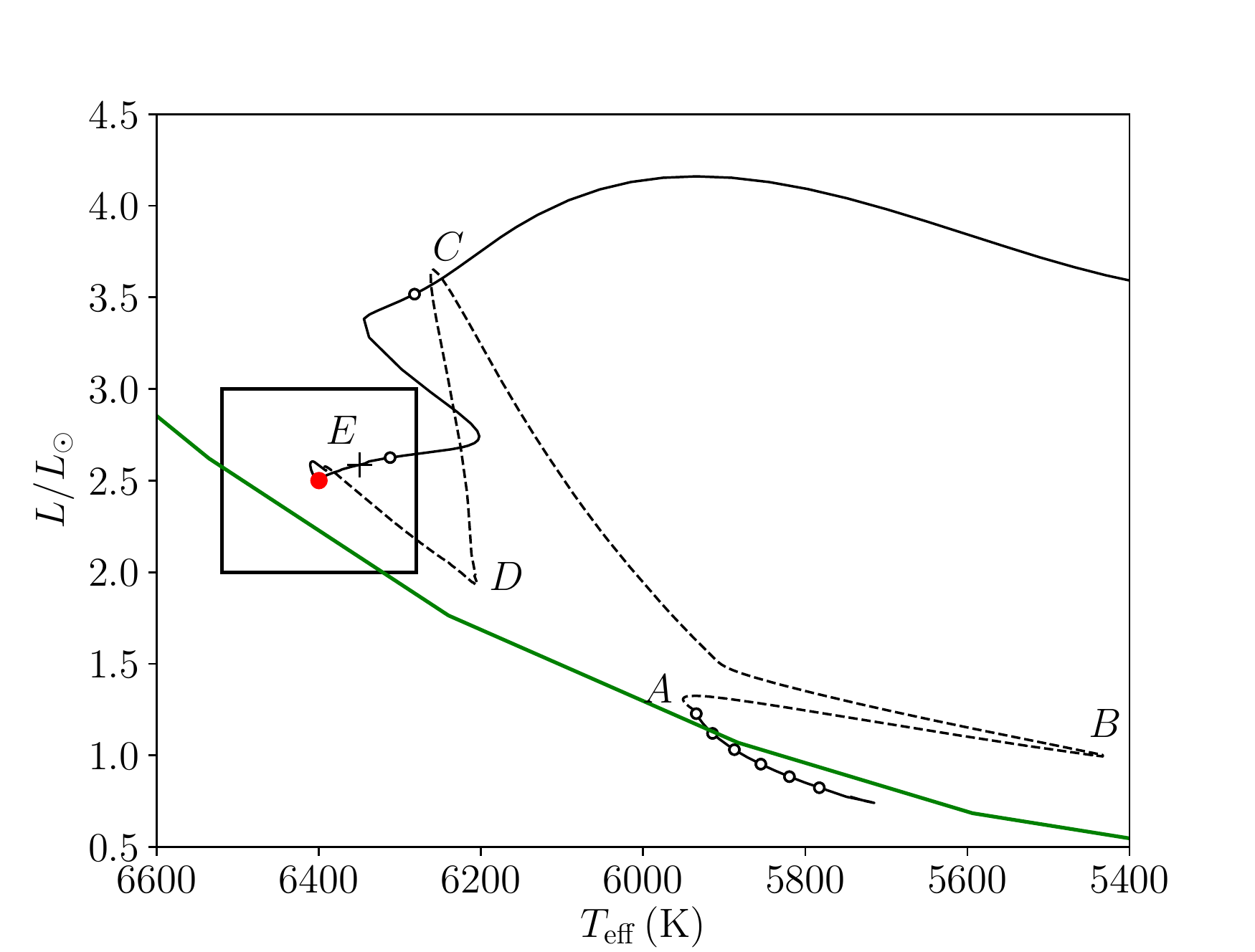}
\caption{The HR diagram ($L$ versus $T_{\rm eff}$) for the evolution of the secondary star into a BSS. The phases with no MT are shown by the solid line. The series of open circles mark 1, 2, ..., 6 Gyrs. The MT phase is shown by the dashed line. The observed $L$ and $T_{\rm eff}$ for WOCS 5379 are shown as the red dot. The box around the red dot indicates the 1-$\sigma$ uncertainty on $T_{\rm eff}$ and the range of luminosity due to brightness variability. 
The best-fit model is shown as a black cross. The green solid line is the ZAMS line from \citep{2015ApJ...814..163G}.}
\label{BSS_HR_best.pdf}
\end{figure}

Figure \ref{BSS_HR_best.pdf} shows the evolutionary track of the secondary star of the Best-Fit Model on the HR diagram, compared to the observed parameters of the BSS. The bolometric luminosity and $T_{\rm eff}$ from the spectroscopic analysis of \cite{2019ApJ...885...45G} are displayed with the red dot. The box indicates the 1-$\sigma$ uncertainty on $T_{\rm eff}$ and the range of luminosity due to brightness variability. 
The BF model is marked with a black cross. The reason why the BF model is not exactly at the red dot is because the WD $T_{\rm eff}$ at the red dot is still too high to explain the data. The letters mark important stages during the evolution. A is the onset of the RLOF. The star reaches a minimum $T_{\rm eff}$ at B. The maximum $L$ during the accretion is at C. Then the star drops down to a local minimum in $L$ at D. The cessation of the MT is marked as point E. The detailed physics causing these evolutionary features are discussed in Section \ref{sec: stg_evol}.

In brief, the secondary star begins as a 1.01 $M_{\odot}$ star on the ZAMS, at the coordinates $(T_{\rm eff}/{\rm K},\, L/L_{\odot})=(5700, 0.75)$. The series of open circles mark 1, 2, ..., 6 Gyrs. The evolution of the secondary star as it accretes during MT is shown by the dashed line. The mass accretion begins at 6.13 Gyr and ends at 6.18 Gyr. During the accretion, the track first evolves to cooler effective temperatures. Then the accretor becomes hotter and more luminous, reaching a local maximum of $L = 3.7~L_{\odot}$ at 6200 K. The luminosity next drops due to a shrinking radius while the $T_{\rm eff}$ changes little. There is a second phase of increasing $L$ and $T_{\rm eff}$ before the MT ceases. At this point the accretor has become a 1.20 $M_{\odot}$ BSS. Its interior structure corresponds to the late MS phase and shortly thereafter begins to move onto the subgiant branch. The simulation is terminated at the accretor's RGB phase.

\begin{figure}[tp]
	\includegraphics[width=0.5\textwidth]{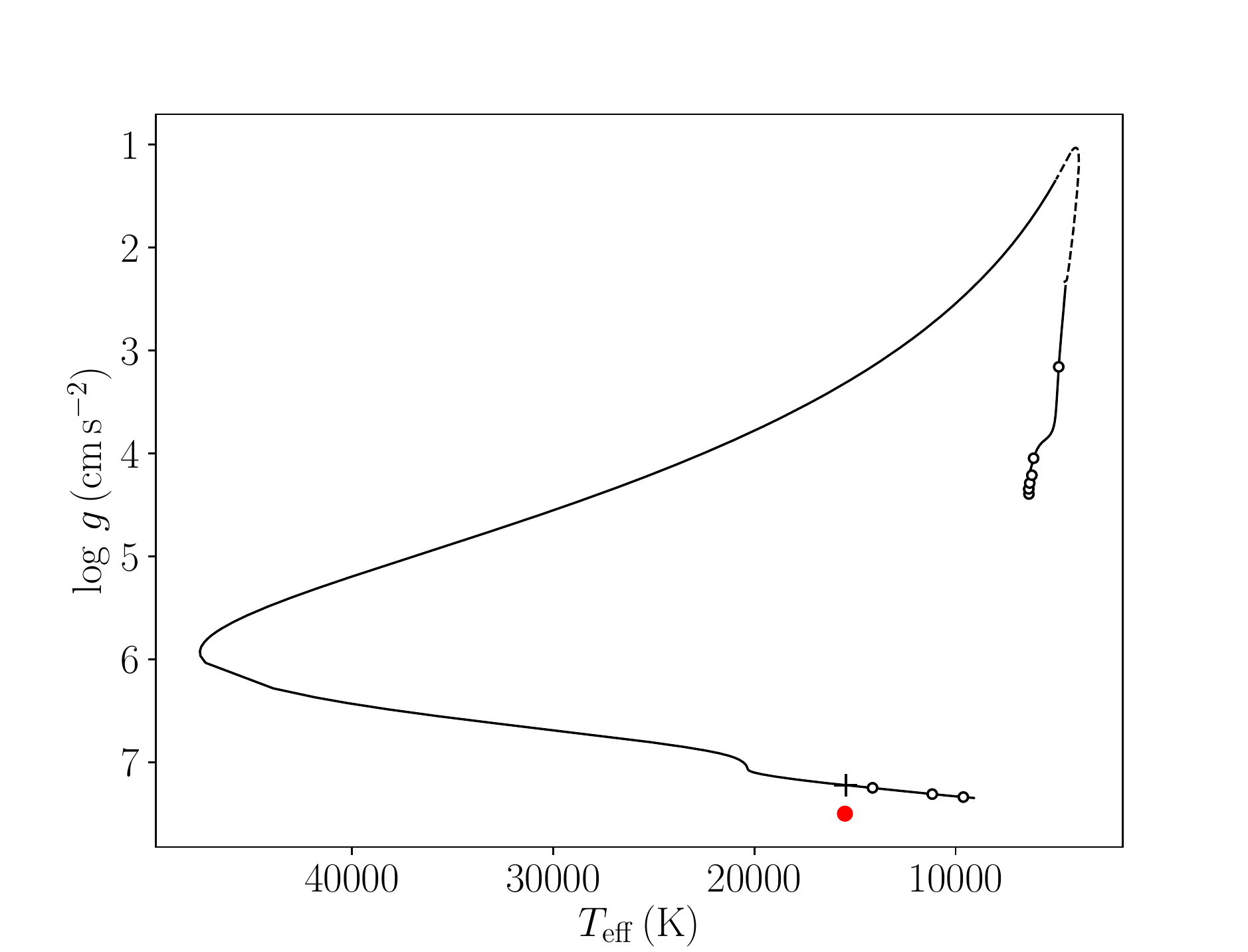}
	\caption{The evolutionary track of the primary star on the $\log g$ - $T_{\rm eff}$ plane. 
The figure follows the description of Figure \ref{BSS_HR_best.pdf}.}
	\label{WD_loggTeff_best.pdf}
\end{figure}

Figure \ref{WD_loggTeff_best.pdf} shows the evolutionary track of the initial primary star, the WD progenitor, in $\log g$ - $T_{\rm eff}$ plane. Again, the open circles mark every 1 Gyr, and the dashed line indicates that MT is occurring. The beginning of the evolution starts from the ZAMS phase of the primary star $(T_{\rm eff},\,\log g)=(6200,4.4)$. After 6 Gyr, the primary star becomes an RGB star and then fills its Roche-lobe radius, launching the MT. After losing most of its hydrogen layer, the donor star shrinks within its Roche-lobe radius and the MT stops. After the MT, the donor star enters the pre-WD phase and the WD cooling phase. During the cooling phase the evolutionary track reaches the observed $T_{\rm eff}$ of the WD. However, the theoretical $\log g$ is smaller than the observed quantity. The mass of the WD in best-fit model is $0.33M_{\odot}$, which is $0.1M_{\odot}$ smaller than the spectroscopic measurement of the WOCS 5379 WD of \citet{2019ApJ...885...45G}.

Lastly, while the BF model is closest to both the binary and BSS properties and the WD $T_{\rm eff}$, other models generated nearby in the initial parameter space also fall within the observational uncertainties. For example, starting with a 0.03 $M_{\odot}$ more massive accretor yields a BSS more luminous than the observation but still inside the measurement uncertainty and still matching the orbital period and WD $T_{\rm eff}$, albeit little change in the WD mass. (See also Sections \ref{BSS_formation_diff_accr} and \ref{LMTE}.) In the following sections, the BF model is used as the specific example for discussing the evolution of the BSS from the MT channel.

\section{General Physical Considerations in the Evolution of the Best-Fit Model}
\label{bi_evol}

\subsection{Overview of Non-conservative Mass Transfer}

In the best-fit model, MT starts as RLOF during the red giant phase of the donor star. A key to having stable MT is that the mass transferred is not conserved within the system (e.g., \citealt{2012ApJ...744...12W}). As the system loses angular momentum with the loss of mass, the binary semi-major $a$ expands depending on the combination of changes in the donor star mass $M_1$, accretor star mass $M_2$, and orbital angular momentum $J$. The orbital angular momentum is

\be
J = M_1 M_2\sqrt{\frac{Ga}{M_1+M_2}},
\ee
 where $G$ is the gravitational constant. The time derivative of the orbital semi-major axis then becomes 
 
 \be
 \frac{\dot{a}}{a} = 2\frac{\dot{J}}{J}+\frac{\dot{M_1}+\dot{M_2}}{M_1+M_2}-2\frac{\dot{M_1}}{M_1}-2\frac{\dot{M_2}}{M_2},
 \ee
where $\dot{M_1}$ is the mass transfer rate of the donor and $\dot{M_2}$ is the mass transfer rate of the accretor. Defining the ratio between $\dot{M_1}$ and $\dot{M_2}$ as

\be
\dot{M_2} = - \beta{\dot{M_1}},
\ee
where the negative sign means the mass loss rate and mass accretion rate have different signs, and the mass ratio between $M_1$ and $M_2$ as

\be
q=\frac{M_1}{M_2},
\ee
then the rate of change in semi-major axis can be written as

\be
\frac{\dot{a}}{a} = \bigg( \frac{(2q+1)(1-\beta)}{1+q}+2\beta-\frac{2}{q}\bigg)\frac{\dot{M_1}}{M_1}.
\label{equ: a_expand_rate}
\ee
Given that the donor star is losing mass, $\dot{M_1}$ is negative. To make $\dot{a}$ positive, so the system can maintain a stable MT even the as donor star expands due to adiabatic mass loss, the quantity in the brackets must also be negative. For any mass ratio greater than unity,
 $\beta<1$ is the key to expand the system during the MT. 
 
For the BF model, the MT efficiency is 22\%. The mechanism of the associated mass loss from the system is unknown and beyond the scope of this paper (but see Section \ref{MT_alpha_mechanism}).

\subsection{The Mass and Age of the WD Companion}

The mass of the product WD, which is slightly greater than the helium core of the donor, is determined in large part by the orbital period at the onset of the MT. Specifically, for a 1.2 $M_{\odot}$ MS star, the core is convective and the envelope is primarily radiative. Once the hydrogen abundance drops below 0.01 in mass fraction near the convective-radiative boundary and helium is the dominant element, the enclosed mass is defined as the helium core.  

\begin{figure}[tp]
	\includegraphics[width=0.5\textwidth]{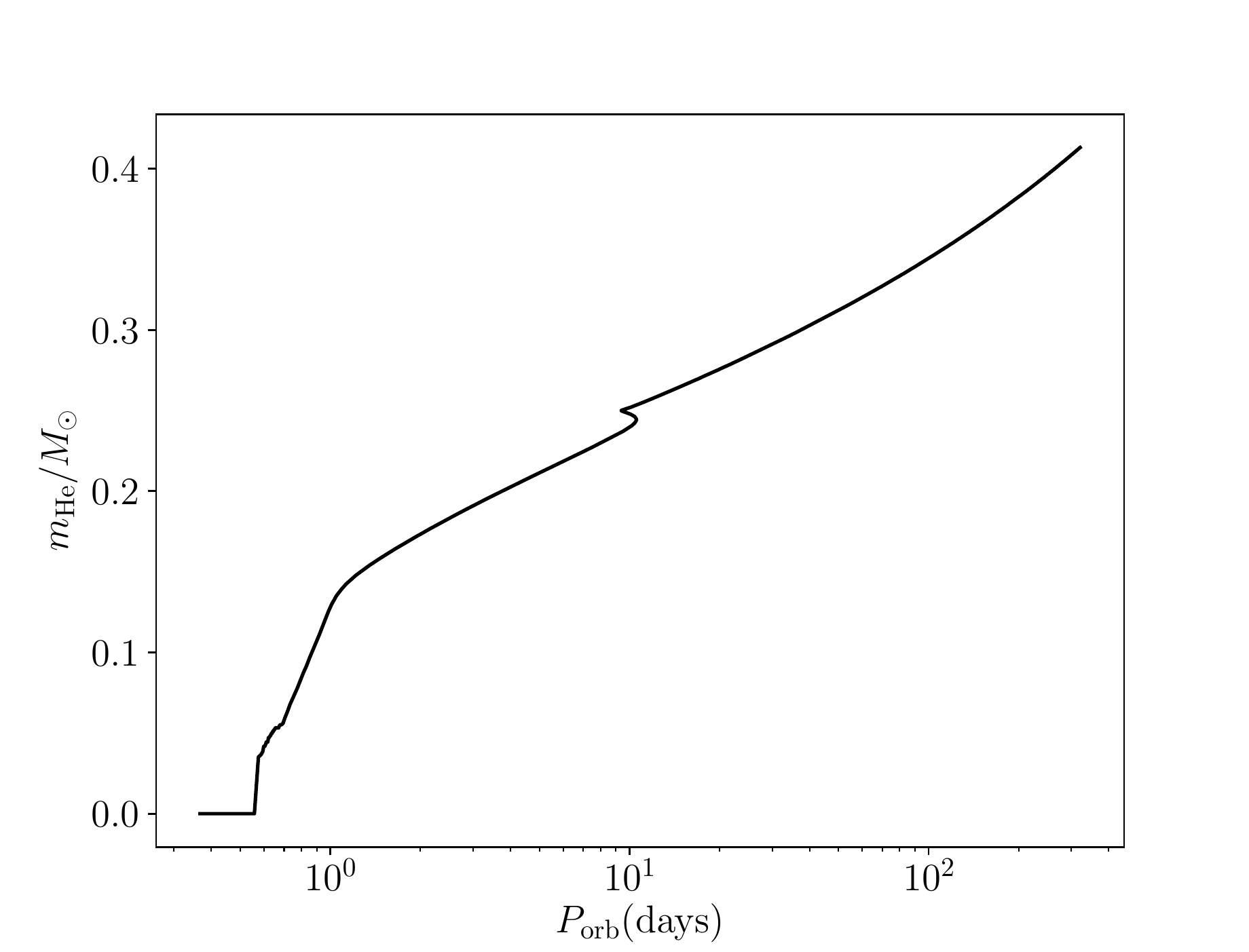}
	\caption{Helium core mass $m_{\rm He}$ for a 1.2 $M_{\odot}$ at the onset of the RLOF as a function of orbital period with a 1.0 $M_{\odot}$ companion star. Based on MESA modeling of 1.2 $M_{\odot}$ single-star evolution.}
	\label{fig:Mhe_Porb_12Msun.pdf}
\end{figure}

The relation between the helium core mass and the system orbital period at the beginning of RLOF can be derived by connecting the stellar radius $R_1$ and the semi-major axis $a$ through MESA's definition of the Roche-lobe radius $R_{\rm RL}$ \citep{1983ApJ...268..368E},

\be\label{RRL_Eggleton}
\frac{R_{\rm RL}}{a}=\frac{0.49q^{2/3}}{0.6q^{2/3}+{\rm ln}(1+q^{1/3})}.
\ee
At the onset of RLOF, the donor star fills its Roche lobe, so $R_1=R_{\rm RL}$. Converting $R_{\rm RL}$ to $P_{\rm orb}$, Figure \ref{fig:Mhe_Porb_12Msun.pdf} shows the helium core mass $m_{\rm He}$ for a 1.2 $M_{\odot}$ star at the beginning of MT. Varying the accretor mass doesn't affect the relation between the $m_{\rm He}$ of the donor star and the $P_{\rm orb}$ at the onset of the RLOF. The radius slightly shrinks and then increases again near $m_{\rm He}=0.25M_{\odot}$ because of the first dredge-up, resulting in a small bump in the middle of Figure \ref{fig:Mhe_Porb_12Msun.pdf}.

If RLOF begins simultaneously with the formation of the helium core, the system orbital period is only 0.56 days. The helium core keeps growing due to the hydrogen shell burning. Notably, for the $P_{\rm orb} = 120$ days of the current (post-MT) WOCS 5379, the helium core mass is 0.35 $M_{\odot}$. This is smaller than the WD mass of 0.42 $M_{\odot}$ measured by Gosnell et al. (2019). The measured WD mass cannot be the core mass when MT starts. 

The measurement uncertainty in the final orbital period is negligible, so resolution needs to be found in the physics of the model. The mass of the BF helium core doesn't grow much during the short MT phase, increasing only from 0.26 to 0.33 $M_{\odot}$.
Although most of the envelope is lost during the MT phase, a thin hydrogen layer keeps burning that increases the helium core mass of the donor, but only slightly. 

Presuming that the RGB interior evolution is correct, within the assumptions of our modeling this difference in the simulation and observed WD masses can only be resolved if the MT starts later and ends later in the RGB evolution. Taking core growth during MT of a few hundredths of solar mass, in order to make a final WD with a mass of $\sim 0.4 M_{\odot}$ the onset of RLOF occurs at a period near 100 days (Figure \ref{fig:Mhe_Porb_12Msun.pdf}). 
Reasonable MT parameters then produce final $P_{\rm orb}$ between 500 - 600 days, far greater than observed for WOCS 5379.

The BF model indicates a WD cooling time of $\sim 593$ Myr, measured by the time between the end of MT to the observed $T_{\rm eff}$ for WOCS 5379 (15500 K).
This cooling time differs from the 250 Myr value of \citet{2019ApJ...885...45G}. The difference arises primarily from (a) element diffusion not being included in our BF WD model; and (b) the WD mass from the BF model being 0.1 $M_{\odot}$ less massive than the \citet{2019ApJ...885...45G} mass, derived from the observed $\log\,g$. For helium core WDs, the general trend is that less massive WDs have longer cooling times \citep{2013A&A...557A..19A}. With MESA modeling, we find that a WD mass difference from 0.42 $M_{\odot}$ to 0.33 $M_{\odot}$ increases the cooling time by $\sim$ 0.1 Gyr.

The decrease in WD cooling time due to element diffusion is discussed in \citet{2016A&A...595A..35I}. For their $Z=0.02$ stars, the inclusion of  diffusion decreases 
cooling times by several hundred Myr.

For WOCS 5379, we choose to use the cooling times of \citet{2013A&A...557A..19A}, as did \citet{2019ApJ...885...45G}. Interpolating in their Table 3 for a 0.33 $M_{\odot}$, we adopt a cooling time to the current temperature of the BF WD of 0.3 Gyr. This shifts the model age for WOCS 5379 to 6.5 Gyr, still well within the current age range for NGC 188. We note that these small shifts in the WD age have little impact on the BF BSS, in fact shifting it slightly closer to the nominal position of WOCS 5379 in the HR diagram (Figure \ref{BSS_HR_best.pdf}).

\subsection{Stability Analysis of the Mass Transfer}

\begin{figure}[tp]
	\includegraphics[width=0.5\textwidth]{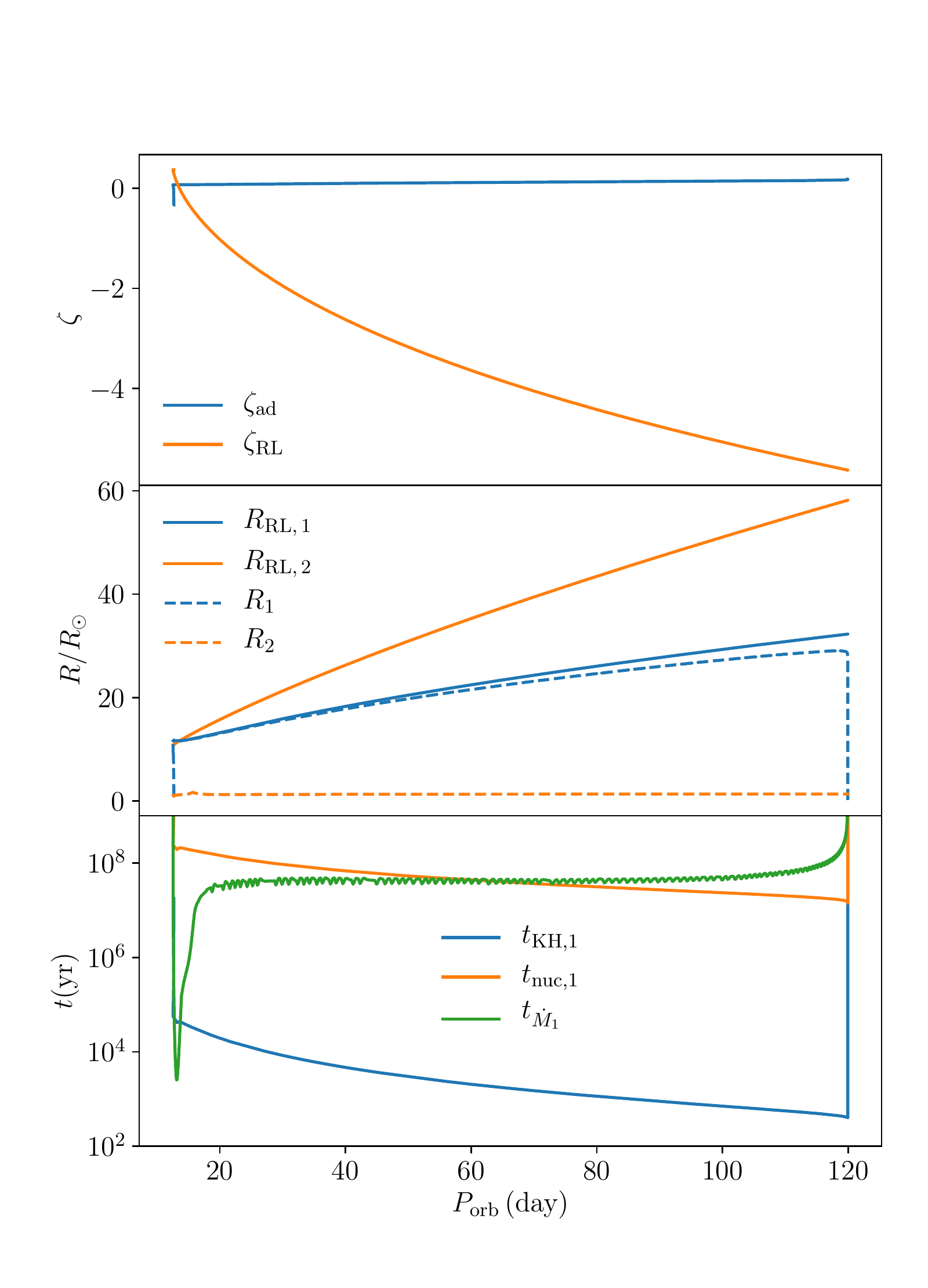}
	\caption{The top panel displays the result of the stability analysis, where $\zeta_{\rm ad}>\zeta_{\rm RL}$ indicates stable MT. The middle subplot shows the Roche-lobe radii $R_{\rm RL,1}$ (blue solid line) and $R_{\rm RL,2}$ (orange solid line) compared to the stellar radii $R_1$ (blue dashed line) and $R_2$ (orange dashed line). The last panel compares $t_{\rm KH}$ (blue), $t_{\rm nuc}$ (orange) and $t_{\dot{M}}$ (green) of the donor star. All are shown in terms of $P_{\rm orb}$.}
	\label{Mdot_P_bi2.pdf}
\end{figure}

The top panel of Figure \ref{Mdot_P_bi2.pdf} compares analytic descriptions of the donor's response $\zeta_{\rm ad}$ and the Roche-lobe radius change $\zeta_{\rm RL}$ during MT, as a function of orbital period. Following \citet{2012ApJ...744...12W}, these two quantities are defined as

\be
\zeta_{\rm ad}\equiv \bigg( \frac{d \log R_1}{d \log M_1} \bigg)_{\rm ad},
\ee

\be
\zeta_{\rm RL}\equiv \frac{d \log r_{\rm RL}}{d \log M_1}.
\ee
Stable MT is defined as $\zeta_{\rm ad} \geq \zeta_{\rm RL}$. To understand this criterion, if both $\zeta$ are negative, $0>\zeta_{\rm ad} \geq \zeta_{\rm RL}$, then $R_1$ expands less than $R_{\rm RL}$ due to the decrease in $M_1$. This will lead to stable MT. Similarly, if the two $\zeta$ are positive, $R_1$ shrinks more than $R_{\rm RL}$ when the donor star loses mass. This also results in stable MT. 

Assuming an $n=3/2$ polytrope model for the stellar structure, the adiabatic response of the donor star can be written as

\be
\begin{split}
\zeta_{\rm ad}=&\frac{2}{3}\frac{m_{\rm He}}{1-m_{\rm He}}-\frac{1}{3}\frac{1-m_{\rm He}}{1+2m_{\rm He}}-0.03m_{\rm He}\\
&+0.2\frac{m_{\rm He}}{1+(1-m_{\rm He})^{-6}} .
\end{split}
\ee

Here $m_{\rm He}$ is the helium core mass fraction of the donor star. The expression for the response of the donor star's Roche lobe is

\be
\zeta_{\rm RL} = \frac{\partial{\rm ln}a}{\partial {\rm ln}M_1}+\frac{\partial{\rm ln}(R_{\rm RL}/a)}{\partial {\rm ln}q}\frac{\partial{\rm ln}q}{\partial {\rm ln}M_1}.
\ee
If the mass loss from the donor star's stellar wind is ignored, the term $\partial{\rm ln}a/ \partial {\rm ln}M_1$ can be written as
\be
 \frac{\partial{\rm ln}a}{\partial {\rm ln}M_1}=\frac{2M_1^2-2M_2^2-M_1 M_2(1-\beta)}{M_2(M_1+M_2)}.
\ee
The expression of the second term $\partial{\rm ln}(r_{\rm RL}/a) / \partial {\rm ln}q$ is derived from the Roche-lobe radius approximation equation \citep{1983ApJ...268..368E,1997A&A...327..620S},
\be
\frac{\partial{\rm ln}(r_{\rm RL}/a)}{\partial {\rm ln}q}=\frac{2}{3}-\frac{q^{1/3}}{3}\frac{1.2q^{1/3}+1/(1+q^{1/3})}{0.6q^{2/3}+{\rm ln}(1+q^{1/3})}.
\ee
The third term is the mass ratio change due to the donor mass loss,
\be
\frac{\partial{\rm ln}q}{\partial {\rm ln}M_1}=1+\beta \frac{M_1}{M_2}.
\ee

The top panel of Figure \ref{Mdot_P_bi2.pdf} shows the MT to be unstable at the beginning ($\zeta_{\rm RL}>\zeta_{\rm ad}$). At the early stage of the evolution when there is no helium core, $\zeta_{\rm ad}$ starts as $-1/3$. As the helium core mass is formed during the late MS and early RGB phase, $\zeta_{\rm ad}$ is near 0.  $\zeta_{\rm RL}$ is positive at the very beginning of the MT, based on the system separation, $M_1$ and $q$ when MT starts. Thus the system is unstable.  $\zeta_{\rm RL}$  then drops below zero and becomes smaller than $\zeta_{\rm ad}$ as the Roche-lobe radius of the donor star and the system separation both increase after the onset of MT. 

The middle panel compares four radii in the system. The Roche-lobe radius of the donor star $R_{\rm RL,1}$ is shown by a blue solid line. This line drops a little at the very beginning of the MT and then expands, which agrees with the sign of $\zeta_{\rm RL}$ (positive, then negative). Near the end of MT, the donor star becomes a pre-WD. Due to the near-complete exhaustion of the envelope mass, its radius shrinks fast near $P_{\rm orb}=120$ days and the MT stops. The Roche-lobe radius of the accretor star $R_{\rm RL,2}$ and $R_2$ are shown as orange solid and dashed lines, respectively. The accretor star never approaches its Roche lobe in the simulation. 

The bottom panel compares three instantaneous timescales, the Kelvin-Helmholtz timescale $t_{\rm KH}$, the nuclear reaction timescale $t_{\rm nuc}$, and the mass-loss timescale of the donor star $t_{\dot{M}_1}$. The BF model has unstable MT at the start, so $t_{\dot{M}_1}$ is much shorter than $t_{\rm KH,1}$ and $t_{\rm nuc,1}$. After the system adjusts itself to stable MT, the thermal adjustment of the donor star drives the MT. At the late phase of the MT, the MT rate becomes small ($\log|\dot{M}_1|=10^{-8}$), so that $t_{\rm nuc,1}<t_{\dot{M}_1}$. Thus at the end of the MT, the helium core  is growing significantly during the MT.

\section{Evolution of WOCS 5379 into a Blue Stragger - White Dwarf Binary}
\label{sec: stg_evol}

This section discusses in detail the evolution of the two stars in WOCS 5379 from an initial ZAMS binary (Table \ref{initial model}) into the current BSS-WD pair (Table \ref{observation vs model}), based on the BF simulation. Recall that the evolutionary tracks of both stars in the theoretical H-R diagram are given in Figures \ref{BSS_HR_best.pdf} and \ref{WD_loggTeff_best.pdf}.

\begin{figure}[tp]
	\includegraphics[width=0.5\textwidth]{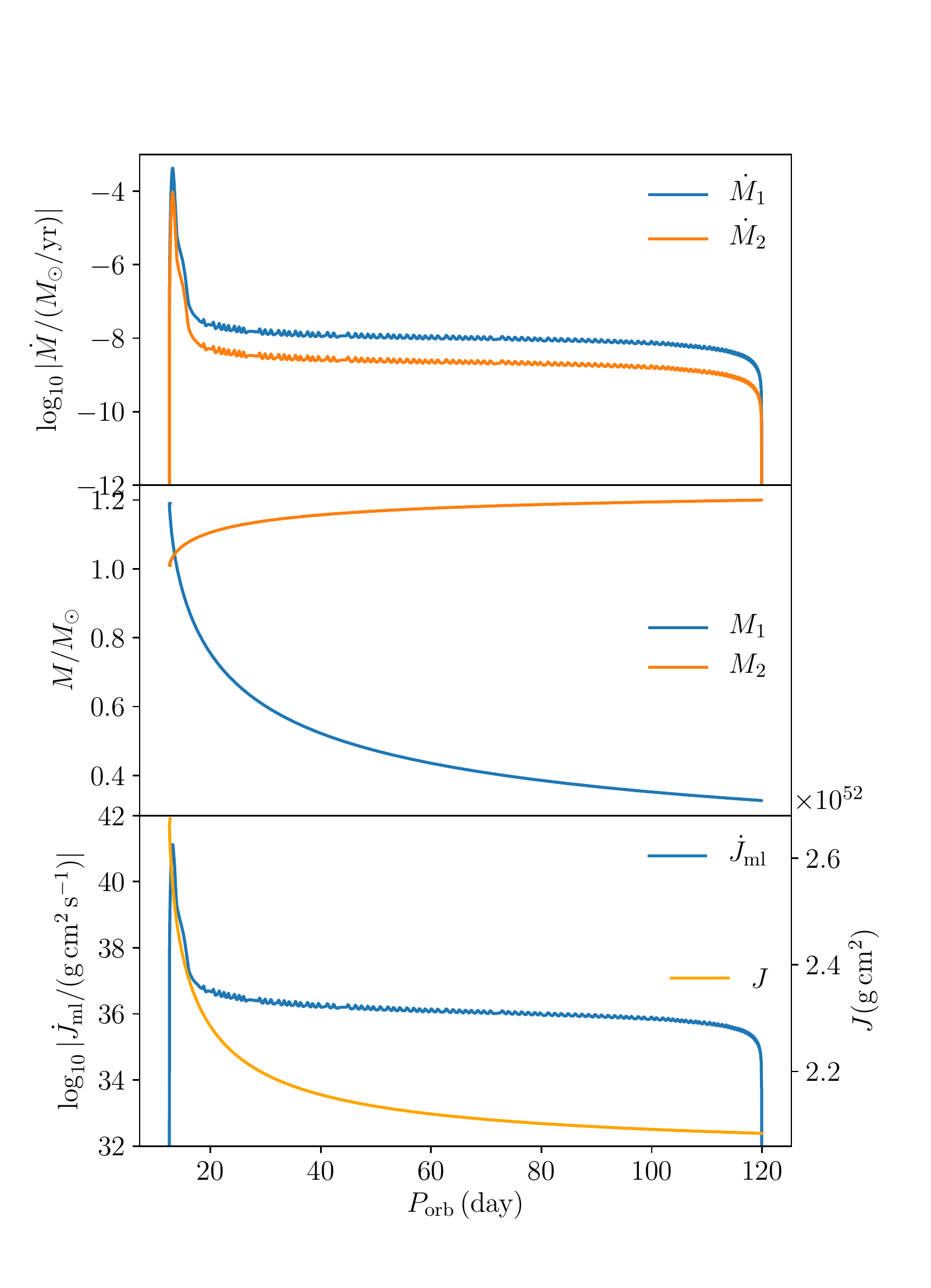}
	\caption{The upper plot shows $\dot{M}$ as a function of $P_{\rm orb}$ for the donor star (blue) and the accretor star (orange), with an initial $P_{\rm orb}=12.7$ days. The middle panel gives the mass of the donor star (blue) and accretor star (orange) during the evolution. The last panel displays the angular momentum loss due to the mass loss rate from the system and the total orbital angular momentum of the system.}
	\label{Mdot_P_bi.pdf}
\end{figure}

Figure \ref{Mdot_P_bi.pdf} displays the system parameters in terms of $P_{\rm orb}$ for the BF model. The evolution begins with an initial $P_{\rm orb}$ of 12.7 days. The system separation doesn't change until MT begins. The MT ceases at $P_{\rm orb}=120$ days, which matches the observed period. The top and middle panels show $\dot{M}$ and stellar mass, respectively, for the donor and accretor stars. The BF model has a MT efficiency of 22\%. Thus, $|\dot{M}_2|$ is always smaller than $\dot{M}_1$, and the total mass $M_1+M_2$ is not a constant during the evolution. 

At the very beginning of MT, the MT is unstable for 7 Myr, with a peak in $\dot{M}$ at $P_{\rm orb}=13.2$ days. During this unstable MT phase, the donor star drops in mass from 1.19 to 0.75 $M_{\odot}$, and the accretor star gains mass from 1.01 to 1.1 $M_{\odot}$. Thus half of the accreted mass occurs during this very short phase of unstable mass transfer. As the separation increases during the MT, the system transitions to stable MT near $P_{\rm orb}=16$ days. When the MT becomes stable, the donor star keeps losing mass from 0.75 to 0.33 $M_{\odot}$, while the accretor mass increases from 1.1 to 1.2 $M_{\odot}$.

The third panel shows the angular momentum loss associated with the mass loss from the system, with the highest loss rate during the unstable MT phase ($10^{41} {\rm g\,cm^2\,s^{-1}}$). The mass loss associated with non-conservative mass transfer is modeled as a fast wind leaving from the vicinity of the accretor. For a given orbital angular momentum $J$, the system loses angular momentum as $J_{\rm ml}$:

\be
\dot{J}_{\rm ml} = J\ \frac{M_1 \dot{M}_1}{M_2(M_1+M_2)}.
\ee

During the unstable MT stage, where $P_{\rm orb}$ increases from 12.7 to 16.0 days, the total $J$ drops from $2.67\times 10^{52}$ to $2.37\times 10^{52}$ ${\rm g\,cm^2}$. During the stable MT phase ($P_{\rm orb}$ changes from 16.0 days to 120 days), $J$ keeps dropping from $2.37\times 10^{52}$ to $2.08\times 10^{52}$ ${\rm g\,cm^2}$. The whole MT phase makes the $J$ decrease by about 22\%, with half of this occurring during the short unstable MT phase.

\begin{figure}[tp]
	\includegraphics[width=0.5\textwidth]{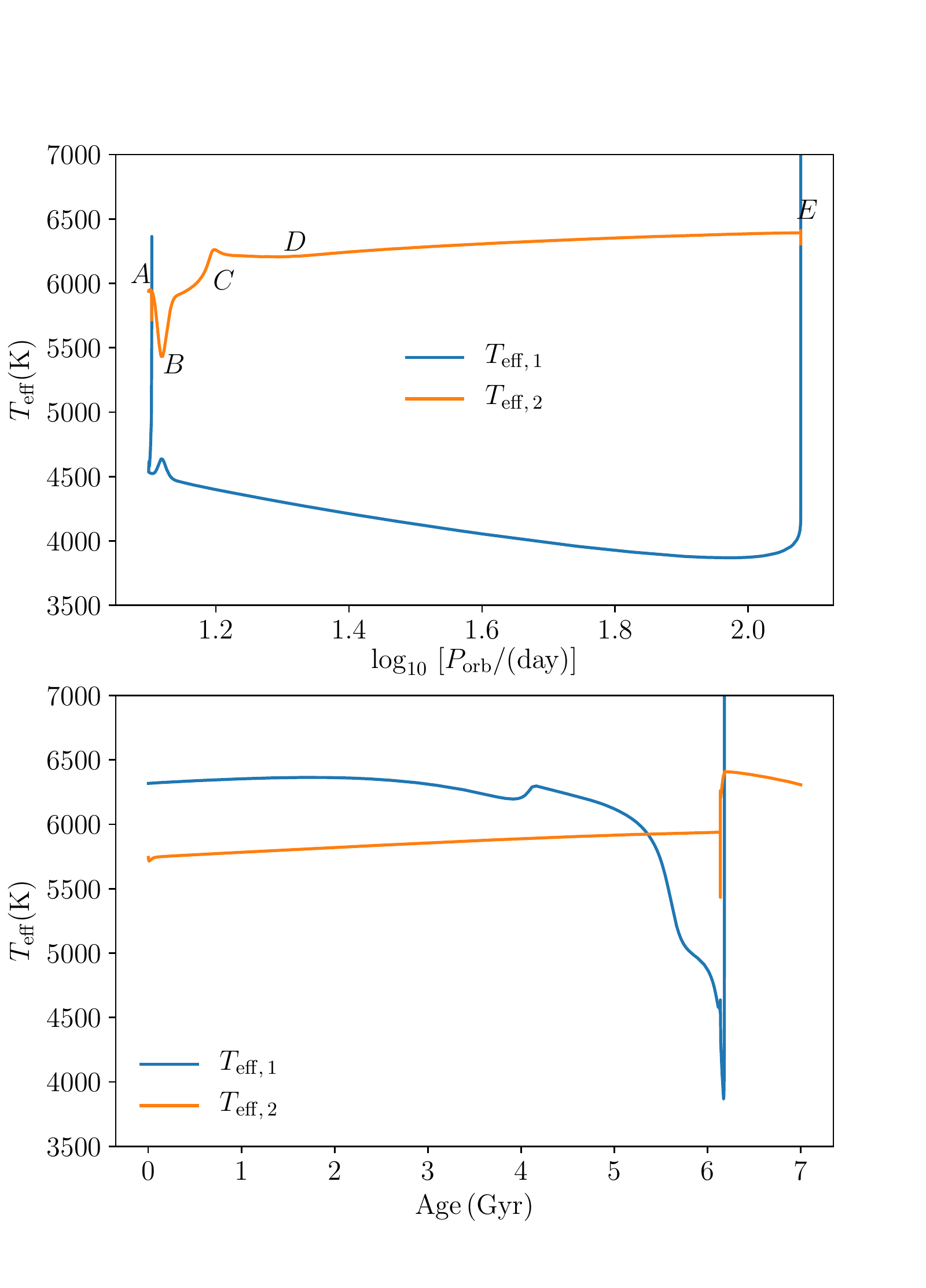}
	\caption{The $T_{\rm eff}$ of the donor star (1) and the accretor star (2) as a function of $P_{\rm orb}$ (top panel) and Age (bottom panel).}
	\label{Teff_compare.pdf}
\end{figure}

\begin{figure}[tp]
	\includegraphics[width=0.5\textwidth]{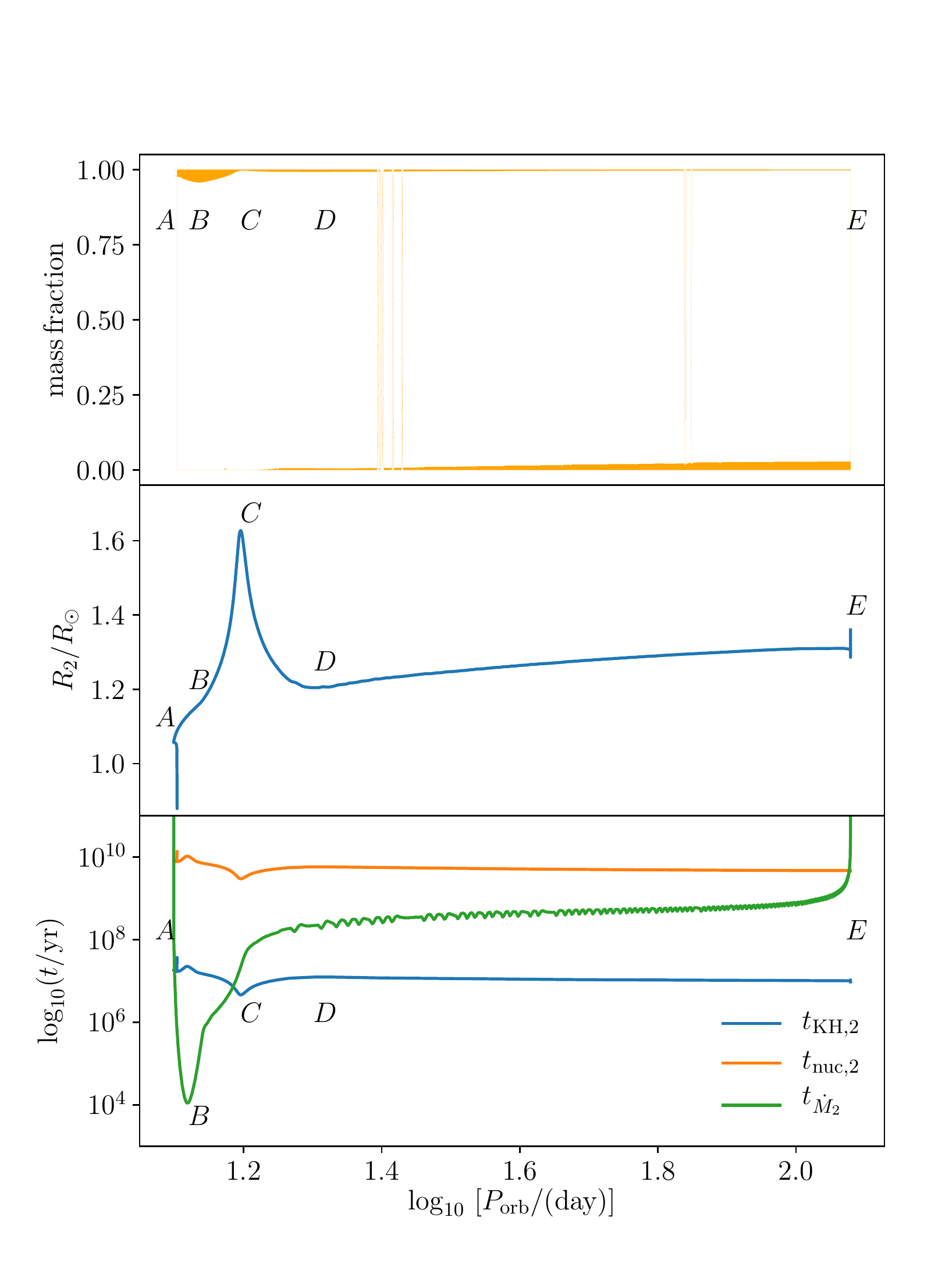}
	\caption{Top: The surface and central convection zone as a function of $P_{\rm orb}$. Between the convection zones is the radiative zone. Middle: $R_2$ in terms of $P_{\rm orb}$. Bottom: Three timescales - thermal $t_{\rm KH,2}$, nuclear $t_{\rm nuc,2}$, and accretion $t_{\dot{M}_2}$ - of the accretor as a function of $P_{\rm orb}$.}
	\label{BSS_kipp.pdf}
\end{figure}

Figure \ref{Teff_compare.pdf} compares the $T_{\rm eff}$ of the initial primary and secondary stars as a function of $P_{\rm orb}$ (top panel) and age (bottom panel). Since the system separation changes very little when MT is not occurring, the top panel reveals more information during the rapid MT phase just after 6 Gyr, while the bottom panel shows better the entire evolution. The letters mark two local minima and one local maximum in $T_{\rm eff,2}$; these are the same evolutionary stages as shown in Figure \ref{BSS_HR_best.pdf} and Figure \ref{BSS_kipp.pdf} with the same letters. Here we analyse Figure \ref{Teff_compare.pdf} and Figure \ref{BSS_kipp.pdf} together to understand the physical quantities (e.g. $L$, $T_{\rm eff}$, $R_2$ etc) during the BSS evolution.

At the beginning of the system evolution, $T_{\rm eff,1}>T_{\rm eff,2}$. As the primary star evolves to the RGB, its radius grows and $T_{\rm eff,1}$ cools down, so at the onset of the RLOF (point A), $T_{\rm eff,1}<T_{\rm eff,2}$. At the beginning of the RLOF, the MT is unstable. From A to B, the accretor star receives mass at a fast rate. This mass augments the existing convective envelope, while the star's inner temperature profile does not change. So the temperature at the radiative-convective boundary does not change, while the thickness of the convective envelope increases in mass and mass fraction. This thicker convective envelope is efficient in lowering the temperature, yielding the lower surface temperature at B. This evolution in the convection zone can be seen in the top panel of Figure \ref{BSS_kipp.pdf}, which shows the convective zone in mass fraction of the BSS. Near B, the surface convective zone is at its greatest depth.
 
At B, the MT rate slows considerably as the MT becomes stable. Also at this time, the size of the surface convective zone shrinks and consequently (Figure \ref{BSS_kipp.pdf})   $T_{\rm eff,2}$ begins to increase. 

From the onset of the RLOF until the secondary reaches a local maximum in $L$ (point C), the radius of the secondary increases (Figure \ref{BSS_kipp.pdf}). With the formation of a central convective zone (Figure \ref{BSS_kipp.pdf}), this growth in radius reverses. At the beginning of the simulation, the core of the 1.01 $M_{\odot}$ accretor is radiative and the surface is convective. Near C, where $M_2=1.08M_{\odot}$, the surface convective zone decreases in mass fraction as the central convective zone appears. This internal transition occurs at the local maximum in $R_2$ (Figure \ref{BSS_kipp.pdf}, second panel), the local maximum in $T_{\rm eff}$ (Figure \ref{Teff_compare.pdf}, top panel), and the maximum $L$ on Figure \ref{BSS_HR_best.pdf}.

From C to D, the accretor continues to acquire material from the donor, but the MT is thermally driven and becomes stable. The bottom panel of Figure \ref{BSS_kipp.pdf} displays the three timescales of the accretor; $t_{\rm KH,2}<t_{\dot{M}_2}$, so the accretor has time adjust to a new thermal equilibrium as the MT is occurring. The middle panel of Figure \ref{BSS_kipp.pdf} shows that the radius drops from C to D. This can also be seen in the dramatic luminosity decrease in Figure \ref{BSS_HR_best.pdf}. After D, the central convective zone becomes the biggest convective zone and continues to grow, so the BSS evolves like a more massive star. From D to E, $R_2$, $T_{\rm eff}$ and $L$ increase slowly as a  function of the orbital period.

As the MT ends, the accretor behaves like an evolved $1.2M_{\odot}$ MS star with a core hydrogen fraction somewhat greater than when the MT began. Shortly thereafter the BSS enters the early RGB phase, near 7 Gyr. Therefore, the radius of the BSS increases again, still at $P_{\rm orb}=120$ days in the second panel. 

The evolution of the donor star is much simpler (Figure \ref{WD_loggTeff_best.pdf}). After the MT ceases, the donor star becomes a pre-WD with a rising temperature from 4700 K to 47600 K, with ongoing hydrogen shell burning. After the hydrogen shell is exhausted, the donor enters the WD cooling phase, dropping in $T_{\rm eff}$ and increasing in $\log\,g$.

\begin{figure}[tp]
	\includegraphics[width=0.5\textwidth]{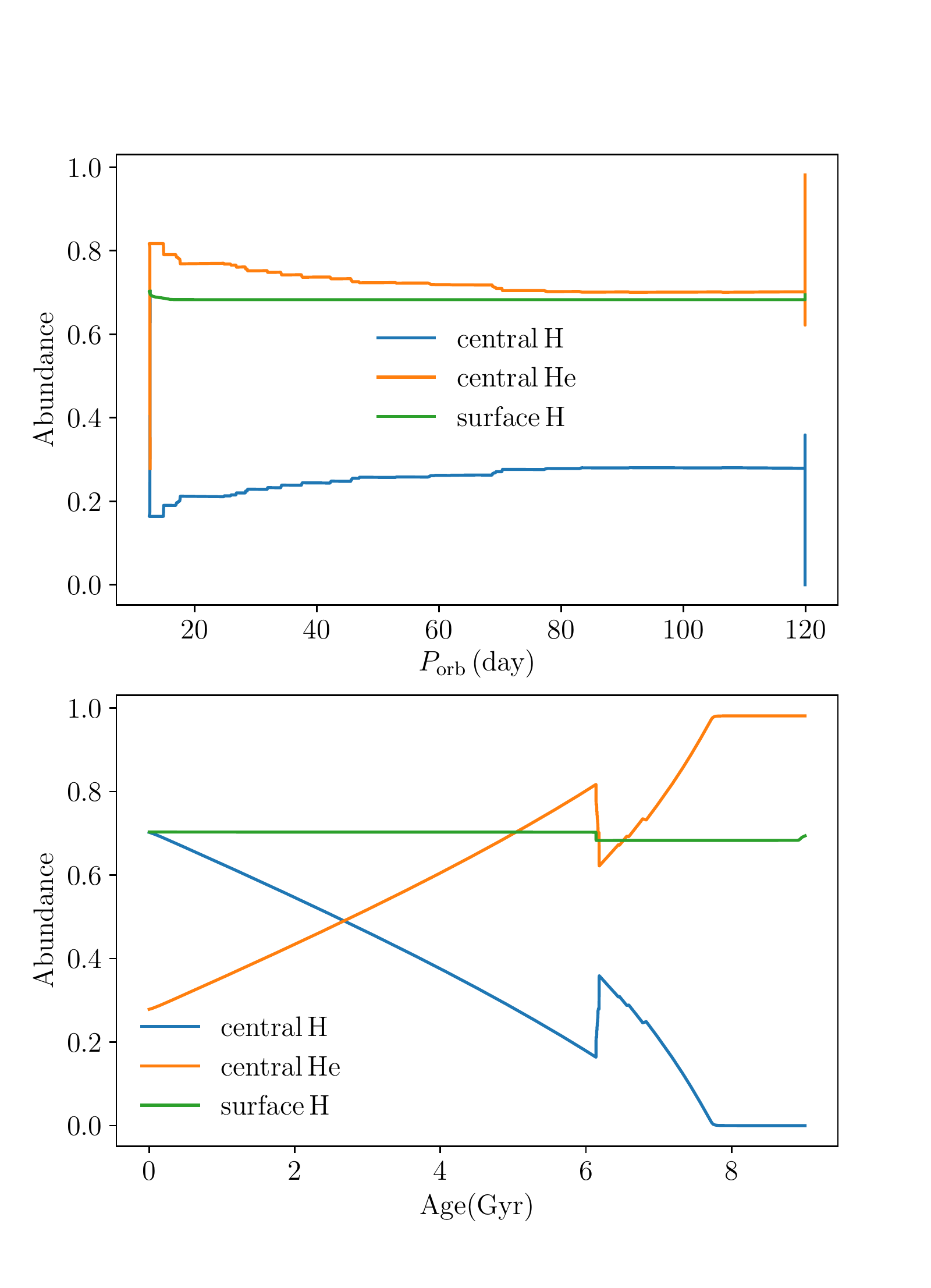}
	\caption{The central (blue), surface (green) hydrogen and central helium (orange) abundance of the accretor as a function of $P_{\rm orb}$ (top) and age (bottom).}
	\label{H_abund_accr.pdf}
\end{figure}

Figure \ref{H_abund_accr.pdf} shows the mass fraction of the central and surface hydrogen in terms of $P_{\rm orb}$ and age. The bottom panel shows that the central hydrogen drops steadily from the ZAMS to 6 Gyr due to single-star MS evolution. The MT phase raises the central hydrogen abundance from 0.16 to 0.36.  The top panel of Figure \ref{H_abund_accr.pdf} shows that this increase is because the central convective zone appears and keeps increasing in size, which allows more hydrogen fuel to mix into the core. Thereafter is a second drop in core hydrogen after the MT, again due to standard single-star MS evolution albeit for a 1.2 $M_{\odot}$ star.  

The surface hydrogen mass fraction is little affected by the MT. The slight drop of the surface hydrogen during the MT is because at the end of MT the donor star moves helium-enriched material onto the accretor.

\section{Discussion: Affects of Physical Parameters}
\label{discussion}

This section discusses the effects on the BSS of changing physically important initial physical parameters of the BF model.

\subsection{BSS Formation via RLOF with Different Accretor Masses}
\label{BSS_formation_diff_accr}

In order to explore the parameter space for making BSS-WD binaries similar to WOCS 5379, Sections \ref{BSS_formation_diff_accr} and \ref{LMTE} discuss changes in the evolution of the BF model with different initial accretor masses and MT efficiencies, respectively, holding all other initial settings the same.

\begin{figure}[tp]
\includegraphics[width=0.5\textwidth]{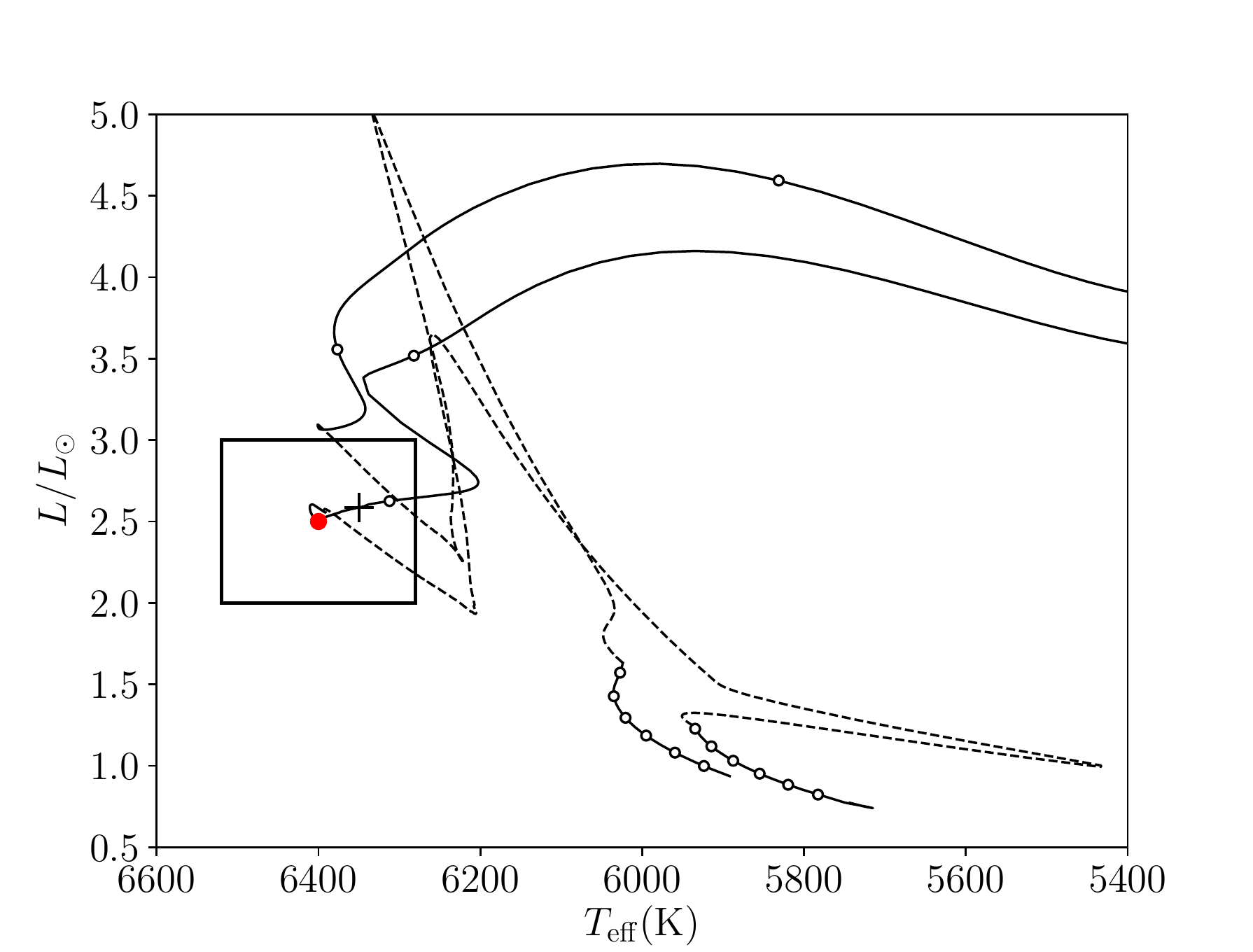}
\caption{The evolutionary tracks forming a BSS from an initial accretor mass of 1.01 $M_{\odot}$(the lower track, the BF model) and of 1.05 $M_{\odot}$ (the more luminous track after the MT). The figure follows the description of Figure \ref{BSS_HR_best.pdf}.}
\label{BSS_diff_accr.pdf}
\end{figure}

Figure \ref{BSS_diff_accr.pdf} compares the evolutionary tracks on the HR diagram with two different accretor masses, 1.01 $M_{\odot}$ (the BF model) and 1.05 $M_{\odot}$. The other settings are the same, specifically the MT efficiency at 22\%, $M_{1,i}=1.19 M_{\odot}$ and $P_{\rm orb,i}=12.7$ days. The track with the 1.05 $M_{\odot}$ accretor starts at $(T_{\rm eff}/{\rm K},L/L_{\odot})=(5900,0.92)$, more luminous and hotter than the start of the BF model. The onset of MT is still unstable for a short time then becomes stable. After MT ends, the BSS has a mass of 1.24 $M_{\odot}$, and while at the same $T_{\rm eff}$ as WOCS 5379, it is more luminous then permitted by the observational uncertainty. The mass change is the same as the BF model while the starting mass is higher than the BF model. So the final product after the MT is more massive and thus more luminous.

For initial accretor masses smaller than 1.01 $M_{\odot}$, it is a challenge to maintain stable MT at the nominal 22\% MT efficiency. Yet higher MT efficiencies result in a lower separation expansion rates during the MT. These lead to unstable mass transfer, since the radius of the donor star now increases faster than its Roche-lobe radius. A common envelope is usually believed form in this situation, after which the final orbital period can be tens of times smaller than the initial separation.

On the other hand, lower MT efficiencies do not produce hot enough products. For a 0.95 $M_{\odot}$ initial accretor mass, the maximum MT efficiency that yields stable MT is 4\%. And the final product after the MT can only reach $(T_{\rm eff}/{\rm K},L/L_{\odot})=(5900,1.3)$, which is far away from the goal.

In summary, higher initial accretor masses with higher MT efficiency can form more luminous BSSs than WOCS 5379. Forming a less luminous but $T_{\rm eff}\sim 6400$ K BSS in NGC 188 challenges the stable RLOF formation channel. 

\subsection{Mass-Transfer Efficiency}
\label{LMTE}

\begin{figure}[tp]
	\includegraphics[width=0.5\textwidth]{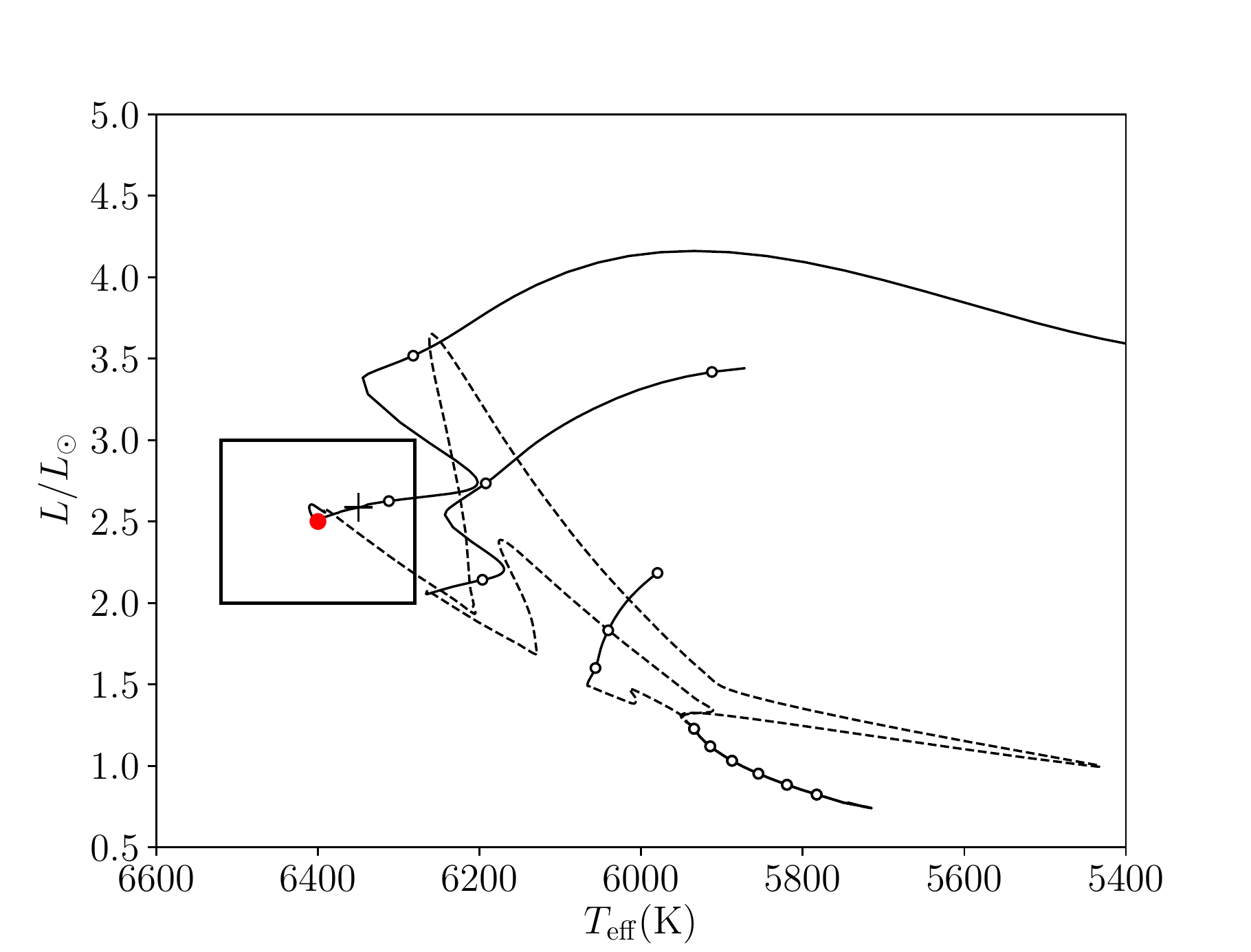}
	\caption{The evolutionary tracks forming a 1.2 $M_{\odot}$ BSS with different MT efficiencies: 5\% (the lowest track after MT), 15\% and 22\% (the highest track after MT, the BF model). The figure follows the description of Figure \ref{BSS_HR_best.pdf}.}
	\label{BSS_diff_beta.pdf}
\end{figure}

Following Section \ref{BSS_formation_diff_accr}, Figure \ref{BSS_diff_beta.pdf} shows a similar experiment, keeping the same initial settings of the BF model but varying the MT efficiencies. Different MT efficiencies only affect the evolutionary tracks after the onset of the RLOF. A MT efficiency of 22\% leads the star to evolve furthest to the red on the HR diagram during MT. On the other hand, with a 5\% MT efficiency there is no accretor evolution to the blue. This occurs because lower MT leads to a less thick convective envelope.

Interestingly, even with a MT of 5\%, the system begins with an unstable MT phase and $\dot{M}> 10^{-6}M_{\odot}/{\rm yr}$. However, less mass is transferred, resulting in a less luminous and colder product near the end of the MT. 

For the 5\% MT efficiency case, the mass of the BSS increases from 1.01 to 1.05 $M_{\odot}$. Thus the BSS does not switch its inner structure from a central radiative zone and an outer convective envelope to a central convective zone and a surface radiative zone. Consequently, there is a smaller decrease in radius during the MT evolution.

\subsection{Forming WOCS 5379 with an Eccentric Orbit}
\label{ecc_chanel}

Currently WOCS 5379 is an eccentric binary with $e=0.24$. Prior to the beginning of MT the binary may have had a tidally circularized orbit. The tidal circularization period of NGC 188 is 14.5 days (Meibom \& Mathieu 2005), longer than the initial binary orbital period of 12.7 days for the BF model above. Thus beginning our simulations with circular orbits is physically plausible.

Even so, the e - log P diagram of NGC 188 does show a couple eccentric binaries with orbital periods around 10 days. Here we begin to explore the impact on the MT and system evolution of an initially eccentric orbit. For specificity we adopt $e=0.24$, which is held constant during the simulation. 

\citet{2019ApJ...872..119H} determine a MT rate for an eccentric orbit with conservative MT, showing that the separation and the eccentricity both change during the MT. The MESA approach is less sophisticated, using an MT rate based on averaging the Roche Lobe radii derived from the changing stellar separations throughout the eccentric orbit, and maintaining a constant eccentricity.

For the case of WOCS 5379, in MESA the large eccentricity further challenges having the stable MT necessary to produce the observed orbital period. The simulation shows that a smaller mass ratio $q$ and lower MT efficiency are necessary to have stable MT.

\begin{figure}[tp]
	\includegraphics[width=0.5\textwidth]{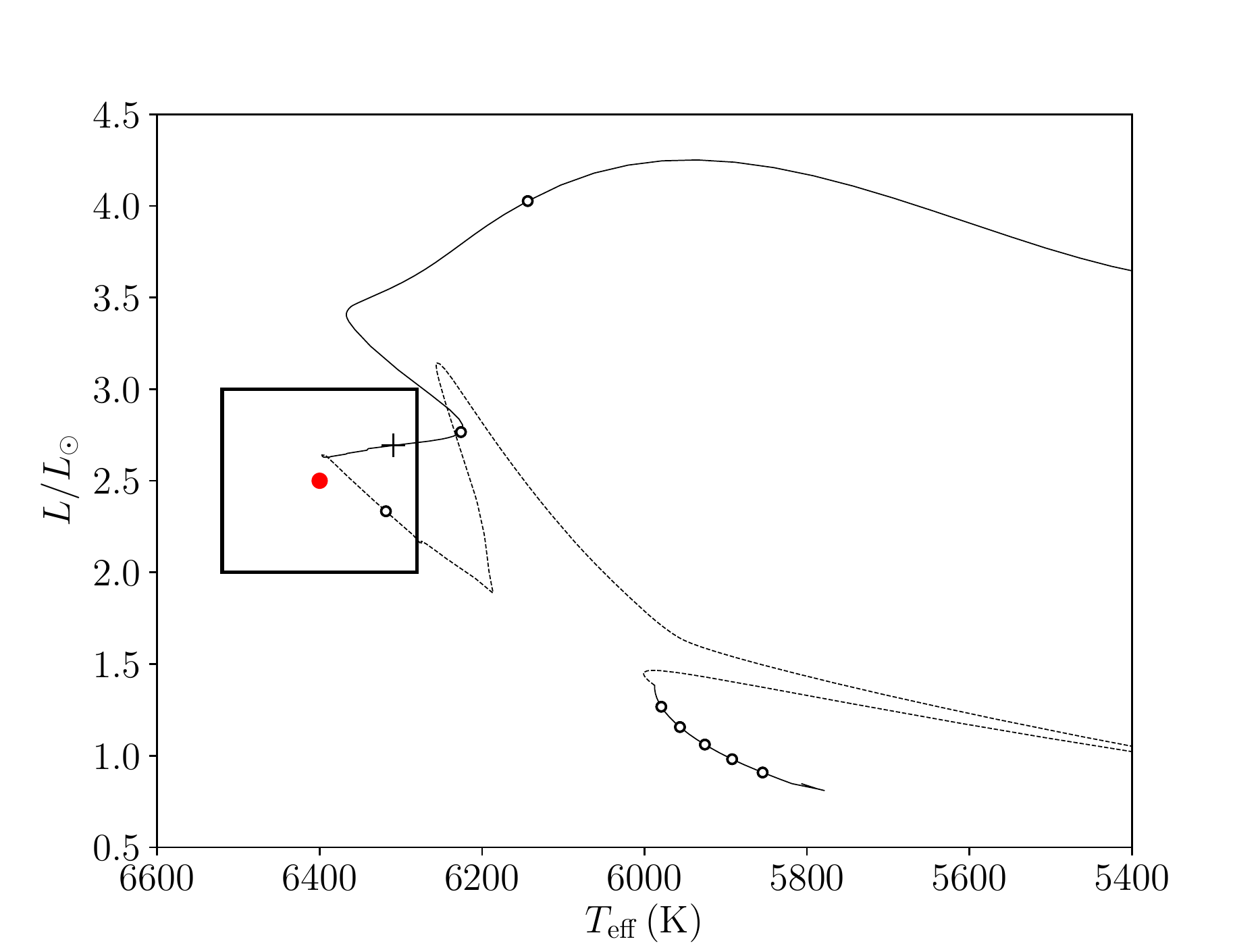}
	\caption{The  evolution of the secondary star beginning in an $e=0.24$ orbit. The best-fit model from the eccentric orbit is shown as a cross. The figure follows the description of Figure \ref{BSS_HR_best.pdf}.}
	\label{BSS_with_ecc.pdf}
\end{figure}

The best-fit model for the eccentric orbit case has $M_1=1.20$, $M_2=1.03$, $P_{\rm orb,i}=10.8$ days and a MT efficiency of 20\%. Again, the best-fit model is taken to be the model that is the closest to the observed data for both the BSS and the WD $T_{\rm eff}$ after MT. Figure \ref{BSS_with_ecc.pdf} shows the evolutionary track of the BSS formed in this model. The differences from the circular-orbit model are not great; the maximum luminosity during the MT phase is smaller.

On the other hand, if the initial binary orbit was in fact circular, the physics of our MESA simulation does not produce orbital eccentricity in the final BSS binary. \citet{2013A&A...556A...4D} discuss how the star spin rate and the eccentricity can affect the duration and the timescale of the MT. \citet{2009ApJ...702.1387S} find that the combination of the MT and tides can either increase or decrease the eccentricity in a close binary system, depending on the mass ratio. However, the separation of the two stars in WOCS 5379 is much greater than the cases they consider. A further detailed study is deserved.

\subsection{Mass Loss from the Vicinity of the Donor}
\label{MT_alpha_mechanism}

Observations of WOCS 5379 are silent on whether the mass lost from the system is from the vicinity of the donor ($\alpha$ mechanism) or of the accretor ($\beta$ mechanism), as we have presumed here. (As defined by \citealt{2006csxs.book..623T}.) In this section, we perform the simulation assuming the mass loss from the system originates in the vicinity of the donor to see if the location of the mass loss affects the formation of the BSS. 

\begin{figure}[tp]
	\includegraphics[width=0.5\textwidth]{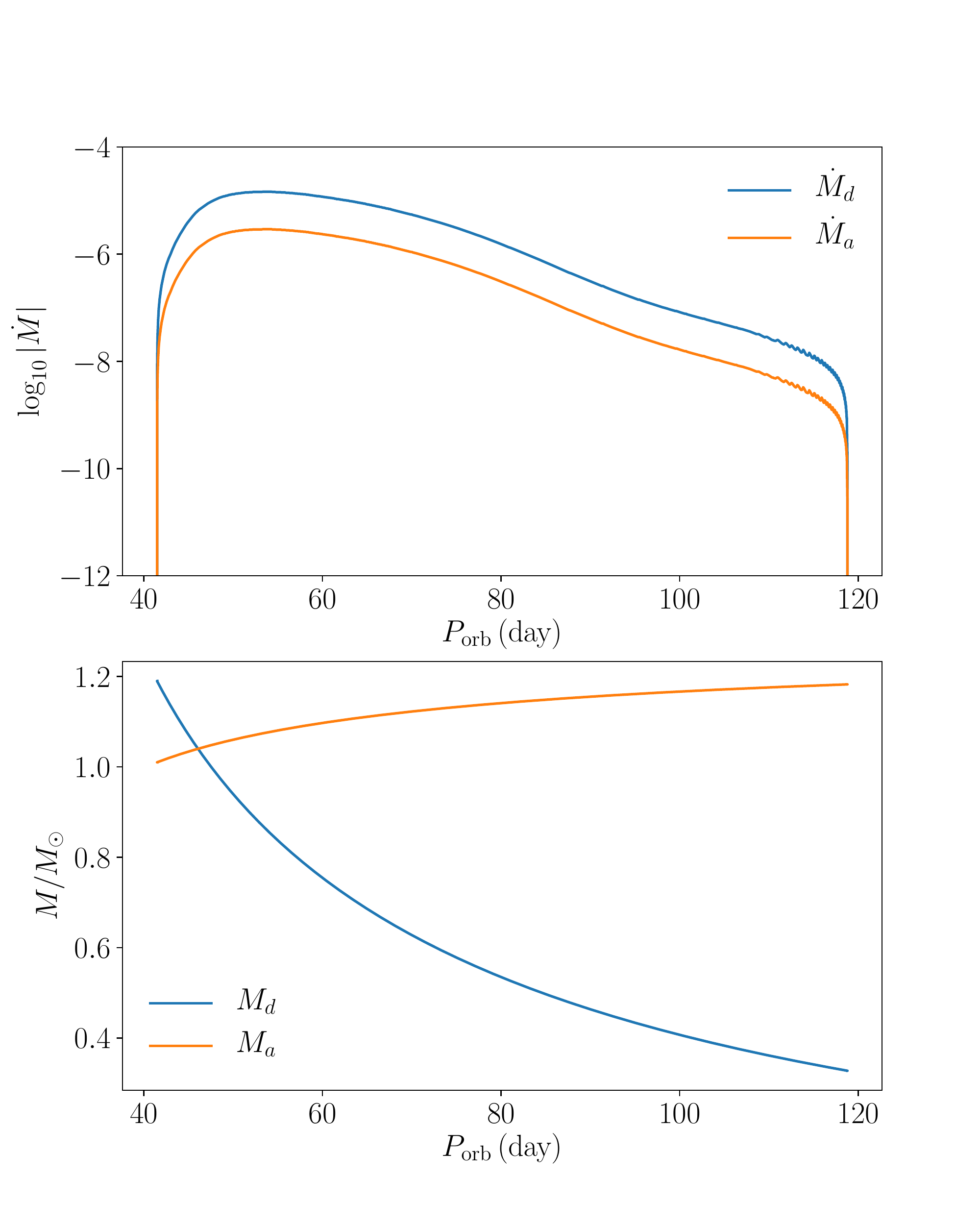}
	\caption{The upper plot shows $\dot{M}$ as a function of $P_{\rm orb}$ for the 1.19 $M_{\odot}$ donor star (blue) and the 1.01 $M_{\odot}$ accretor star (orange) with an initial $P_{\rm orb}=41.5$ days. The bottom panel gives the mass of the donor star (blue) and accretor star (orange) during the evolution.}
	\label{Mdot_P_alfa.pdf}
\end{figure}

Figure \ref{Mdot_P_alfa.pdf} shows the resulting MT rate. The best-fit initial donor and accretor masses and the final BSS and WD masses are very close to the original BF model with the $\beta$ mechanism. However, in order to have the final orbit be $P_{\rm orb}=120$ days, the initial $P_{\rm orb}$ for this model is 41.5 days, as the orbital expansion during the MT is smaller than with the $\beta$ mechanism. The resulting WD mass is again 0.33 $M_{\odot}$. 

This change in initial orbital period can be derived by considering the angular momentum change in losing material from the donor star ${\rm d} J_{\alpha}/J=({\rm d} M/M)(M_2/M_1)$ and from the accretor star ${\rm d} J_{\beta}/J=({\rm d} M/M)(M_1/M_2)$, where ${\rm d} M$ is the mass loss of the total system and $M=M_1+M_2$. For our case, $M_1>M_2$ is always true at the onset of the RLOF. Thus, ${\rm d} J_{\alpha}/{\rm d}M < {\rm d} J_{\beta}/{\rm d}M$ at the beginning of the RLOF, and ${\rm d} J_{\alpha}/{\rm d}M >  {\rm d} J_{\beta}/{\rm d} M$ after $M_1=M_2$. Less ${\rm d} J/{\rm d} M$ results in greater expansion of the orbit. 

For both mechanisms, with the same MT efficiency, $M_1=M_2$ happens at 1.04 $M_{\odot}$. Most of the system mass loss occurs after $M_1 = M_2$. 
Since for the $\beta$ mechanism ${\rm d} J_{\alpha}/{\rm d}M > {\rm d} J_{\beta}/{\rm d} M$ after $M_1=M_2$, the average ${\rm d} J_{\beta}/{\rm d} M$ is smaller ${\rm d} J_{\alpha}/{\rm d}M$. So the system expands more with the  $\beta$ mechanism than the $\alpha$ mechanism. 


Before $M_1=M_2$ at $P_{\rm orb}= 46$ days, the orbital expansion of the $\alpha$ mechanism is more than the $\beta$ mechanism. Consequently, $\zeta_{\rm ad}>\zeta_{\rm RL}$, and the system evolution does not begin with unstable MT (i.e., no spike in $\dot{M}$ in the top panel of Figure \ref{Mdot_P_alfa.pdf}). After $M_1=M_2$ (see Figure \ref{Mdot_P_alfa.pdf}, bottom panel), the system maintains ${\rm d} J_{\alpha}/{\rm d} M$ $>$ ${\rm d} J_{\beta}/{\rm d} M$, so the $\alpha$ mechanism system expansion is less than the $\beta$ mechanism overall.

\begin{figure}[tp]
	\includegraphics[width=0.5\textwidth]{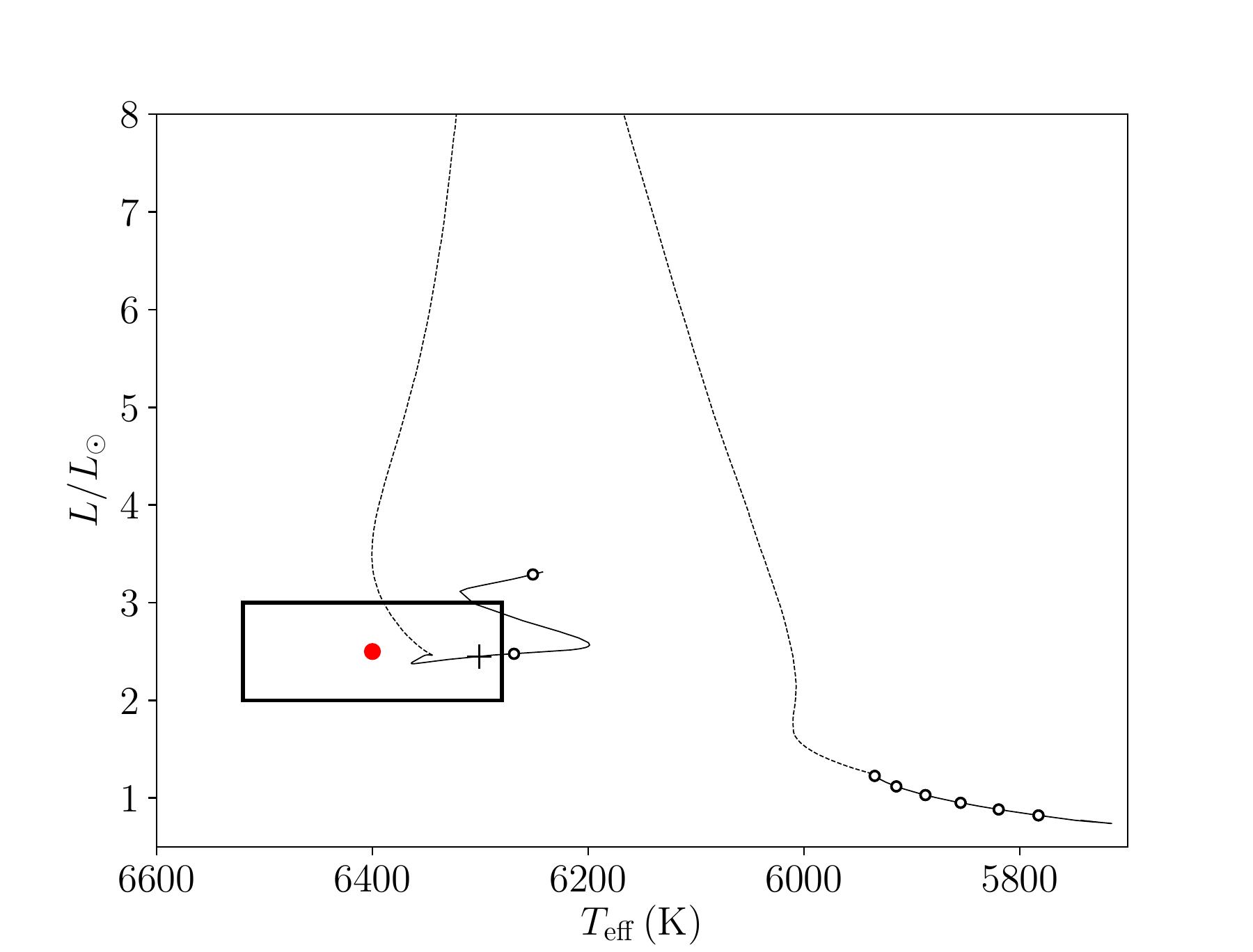}
	\caption{The evolution of the secondary star assuming the mass loss is from the vicinity of the donor star. The best-fit model from the $\alpha$ mechanism is displayed as the cross. The figure follows the description of Figure \ref{BSS_HR_best.pdf}.}
	\label{BSS_MTalpha.pdf}
\end{figure}

\begin{figure}[tp]
	\includegraphics[width=0.5\textwidth]{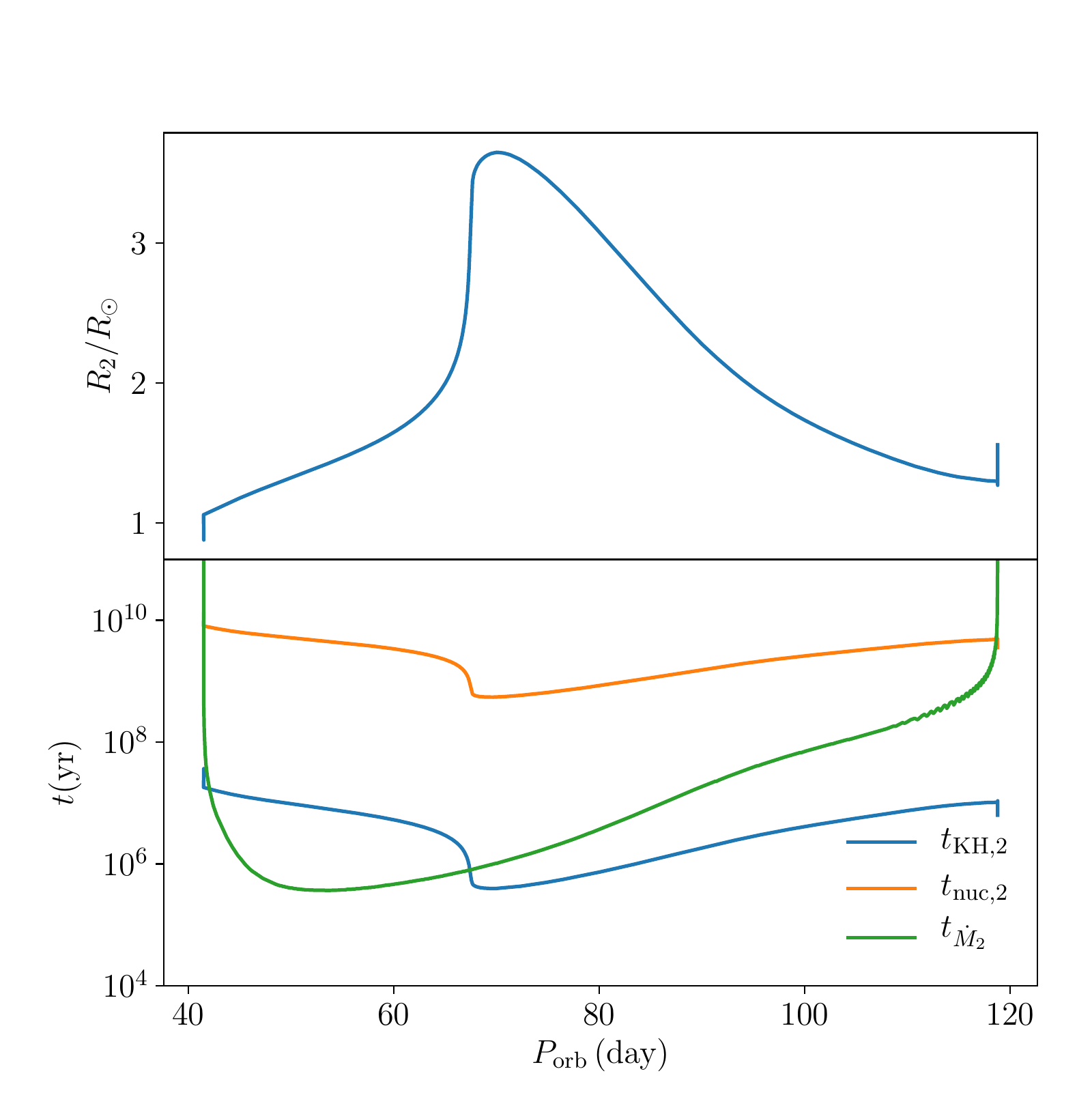}
	\caption{Top: $R_2$ in terms of $P_{\rm orb}$ for the $\alpha$ mechanism. Bottom: Three timescales - thermal $t_{\rm KH,2}$, nuclear $t_{\rm nuc,2}$, and accretion $t_{\dot{M}_2}$ - of the accretor as a function of $P_{\rm orb}$.}
	\label{alpha_R_Porb_accr.pdf}
\end{figure}

The HR diagram in Figure \ref{BSS_MTalpha.pdf} shows the best-fit $\alpha$ mechanism model, which passes within the measurement uncertainty box of WOCS 5379. This model has initial parameters: $M_{d,i}=1.19$, $M_{2,i}=1.01$, $P_{\rm orb,i}=41.5$ with MT efficiency of 20\%. The MT starts after 6.18 Gyr and lasts for about 9.21 Myrs. However, the radius evolution of the accretor is strikingly different. With the $\alpha$ mechanism, the maximum accretor radius is 3.60 $R_{\odot}$ (Figure \ref{alpha_R_Porb_accr.pdf}), while for the $\beta$ mechanism it is 1.63 $R_{\odot}$. Consequently, the maximum luminosity of the accretor during the MT phase with the $\alpha$ mechanism is much greater than with the $\beta$ mechanism. After the donor star becomes a WD, the MT stops with the BSS in satisfactory agreement with the observed WOCS 5379.

The difference in the maximum BSS radius lies in differing physical timescales. The lower panel of Figure \ref{alpha_R_Porb_accr.pdf} shows the three timescales of the accretor during the MT with the $\alpha$  mechanism. At the beginning of the MT, the mass accretion rate for the accretor is $10^{-6}M_{\odot}/{\rm yr}$. At this stage, $t_{\dot{M}_2}$ is shorter than the thermal time. When the MT rate decreases and the thermal time becomes shorter, then the radius of the star responds to the MT.
Compared to the $\beta$ mechanism (Figure \ref{BSS_kipp.pdf}), the thermal timescale MT starts later after the onset of RLOF, which permits the accretor star to grow to such a large radius. Therefore, the maximum luminosity of the BSS for the $\alpha$ mechanism is greater than the $\beta$ mechanism.

\subsection{Stellar Rotation}
\label{sebsec:Rotation}
 Analysing the rotation periods for 12 BSS/MS binaries containing a WD, \citealt{2018ApJ...869L..29L} show that very young post-MT stars are rapidly rotating, with rotation periods as short as 0.4 day. The BSS in WOCS 5379 is not currently an ultra-fast rotator.
\citealt{2018ApJ...869L..29L} give a lower limit on its rotation period of 2.5 days. 

In order to compare the rotation angular momentum and the orbital angular momentum in the BF model, we consider an initial rotation period after MT of 0.4 days. Even at such a short period, the BSS spin angular momentum is much smaller than the system orbital angular momentum (the ratio between them is $10^{-3}$). Thus adding rotation into the simulation will not affect the orbital evolution of the system. Additionally, as the spin energy of the star is a small fraction of its gravitational energy, the rotation will not have a significant impact on the stellar hydrostatic equilibrium.

However rotational mixing can play an important role in stellar evolution. For instance, rotational mixing can alter the amount of hydrogen in a stellar core.
(\citealt{2005MNRAS.358..716S} show this effect in post-collision BSSs in globular clusters.) Adding such rotation mixing in binary evolution models is a goal of our future binary evolution modeling.

\section{Discussion: The Interior Structure and ``Age" of a Blue Straggler}
\label{age_BS}

This paper, and in particular Section \ref{sec: stg_evol}, has presented a detailed story of the formation of an actual BSS through RLOF MT. The details of that formation story are important and fascinating. However, when the story is over, the interior structure of the BSS is the same as that of a normal single star at the same location in the HR diagram, a notable example of the Vogt-Russell Theorem.

\begin{figure}[tp]
	\includegraphics[width=0.5\textwidth]{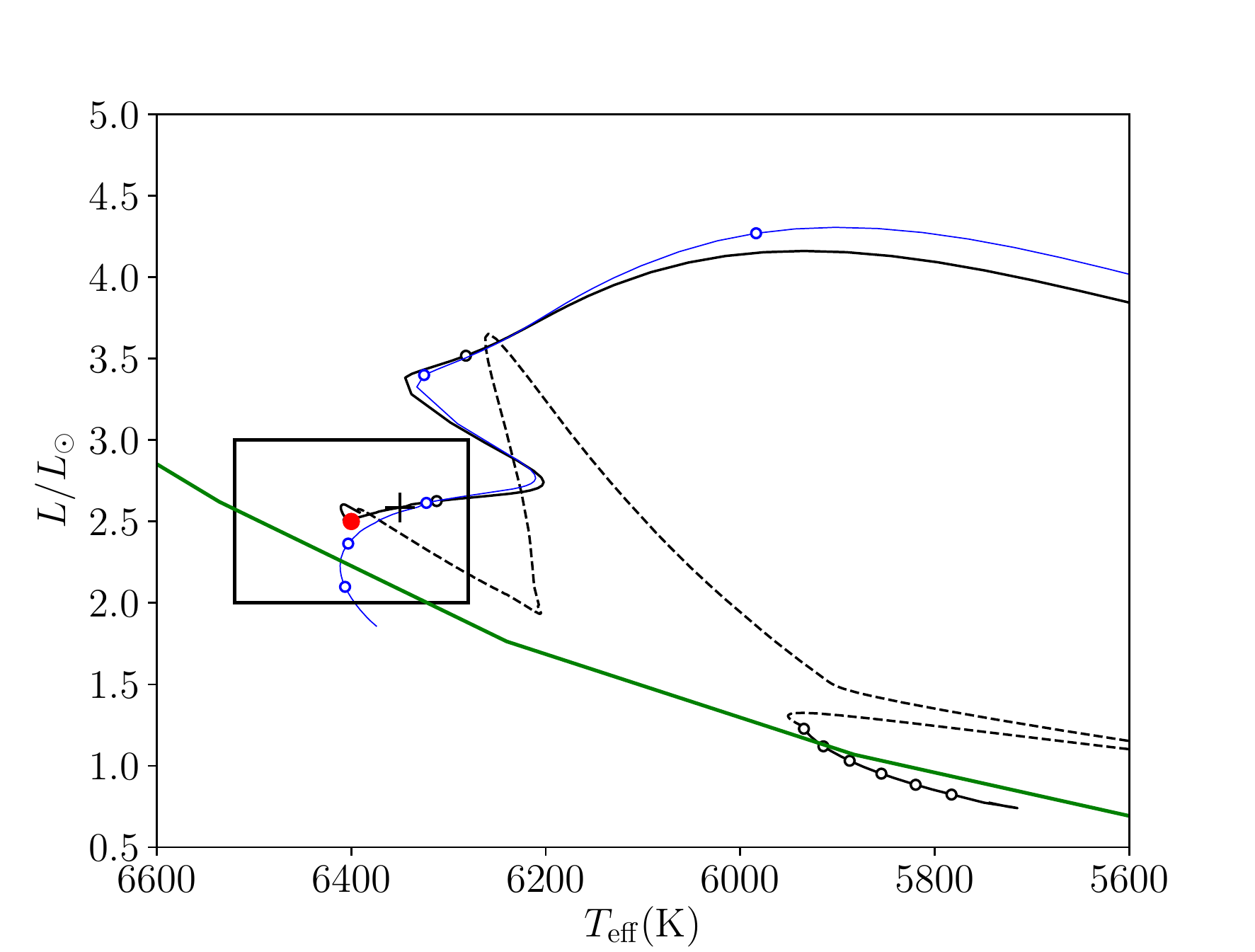}
	\caption{The evolutionary track of the BSS (in black) compared to the evolution of a 1.21 $M_{\odot}$ single star (in blue).  The figure follows the description of Figure \ref{BSS_HR_best.pdf}.}
	\label{BSS_HR_comp.pdf}
\end{figure}

\begin{figure}[tp]
	\includegraphics[width=0.5\textwidth]{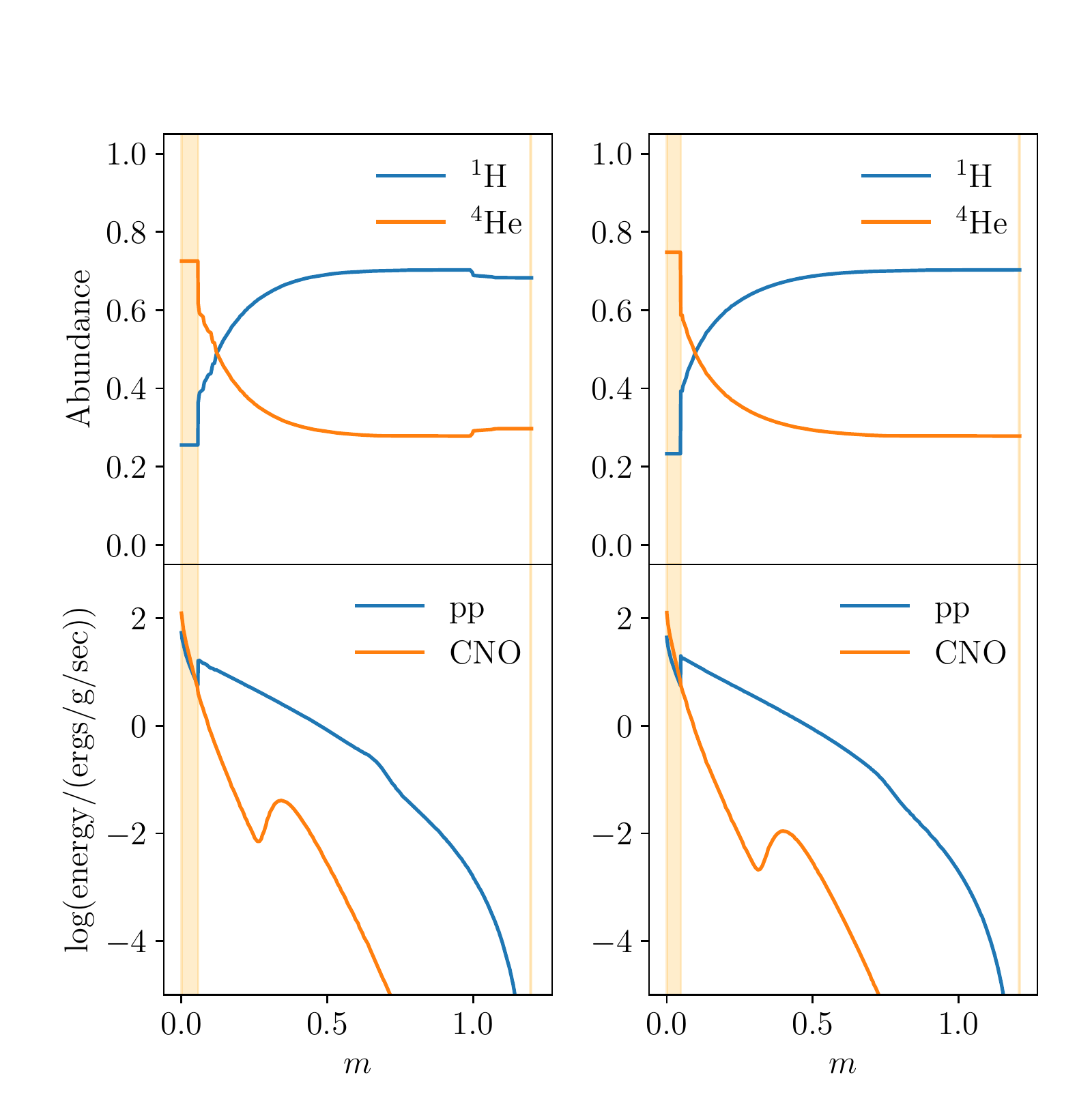}
	\caption{Interior profiles from the BSS BF model (left) and a 1.21 $M_{\odot}$ star (right). The 1.21 $M_{\odot}$ star profile is selected from a single-star evolution track at the closest point to the BF model on the HR diagram. The top and bottom plots show H and He abundances and nuclear energy generation rates, respectively. Convection zones are shaded.}
	\label{profile_comp.pdf}
\end{figure}

In Figure \ref{BSS_HR_comp.pdf} we show the evolutionary tracks of our BF model for WOCS 5379 and of a 1.21 $M_{\odot}$ single star. Both pass through the observed location of WOCS 5379. In Figure \ref{profile_comp.pdf} we show, for both stars, their interior distributions of H and He abundances, fusion energy sources, and convection zones. Despite dramatically different prior evolution histories, the interior structures are essentially indistinguishable.

BSSs are occasionally described as ``rejuvenated" stars, resulting from the addition of mass to a main sequence star by one of several of processes. It is of some interest to consider the age of a BSS in the specific case of WOCS 5379. 

Barring dynamical exchanges, the current binary system WOCS 5379 formed with the cluster NGC 188 and has the same age, e.g. 6.7 Gyr. This is also the \textbf{formation age} of the binary's initial secondary star, now the BSS.

On the other hand, based on the cooling age of the current WD companion, the MT that transformed the secondary star into the current BSS ended only 300 Myr ago, which we call the \textbf{transformation age}. The initial secondary star truly found a fountain of youth in its companion, the initial primary star.

Crucially, the BSS is not a re-born ZAMS star. Its interior structure is that of a 1.21 $M_{\odot}$ MS star with an \textbf{evolution age} of 2.1 Gyr. This evolution age is younger than amost every other star in the current cluster. 

But even so, the BSS is not a rejuvenated 1.01 $M_{\odot}$ star, the original mass of the secondary. The MT has produced a quite new star with an interior structure distinct from a 1.01 $M_{\odot}$ star of any age.

There is one sense the initial secondary star might be considered rejuvenated by the MT. The central H abundance increases from its depleted value before MT of 16\% to 36\% after mass transfer (Figure \ref{H_abund_accr.pdf}). Interestingly, this change is entirely due to structural changes in the star with the addition of mass. With the consequent increase in central temperature and the shift to CNO burning, the central core shifts from radiative to convective, thereby bringing in more fuel. Similar mass transfer onto a lower mass star might not result in a rejuvenation of the central H abundance.

What is unambiguous is that WOCS 5379 will have a much shorter remaining life than its 1.01 $M_{\odot}$ progenitor. We show the future evolutionary track of the BSS in Figure \ref{BSS_HR_comp.pdf}. It will reach the base of the RGB in 9.0 Gyr, at which time its progenitor will still be at $(T_{\rm eff},\,L) = (5900,\,1.75)$ in the late MS stage.

\section{Conclusion}
\label{conclusion}

WOCS 5379 is a blue straggler - white dwarf binary in the open cluster NGC 188. The binary and the cluster are well observed so that the current state of WOCS 5379 is unusually well defined. In this paper we have done a detailed modeling of the evolution of WOCS 5379 from a main-sequence binary to the current system through a mass-transfer event, using the MESA binary evolution module.

After searching across four initial parameters - the primary and secondary masses, the orbital period of the progenitor binary and the MT efficiency - we find a MT model that matches key observed parameters - the BSS $L$ and $T_{\rm eff}$, the WD $T_{\rm eff}$, the current binary $P_{\rm orb}$, and the system age. This model is used to discuss the detailed physics during the evolution. 

In brief, the progenitor binary pairs a 1.19 $M_{\odot}$ star and a 1.01 $M_{\odot}$ star in a 12.7 day circular orbit. During the RGB phase, the primary star fills its Roche-lobe radius and launches MT onto the secondary star. 22\% of the material is accreted and the rest escapes from the system. The mass loss from the system reduces the system angular momentum, and the binary orbit expands. The progenitor secondary star becomes the 1.20 $M_{\odot}$ BS, having evolved to a more luminous (2.5 $
L_{\odot}$) and hotter (6400 K) position in the HR diagram. The core of the primary star becomes the WD companion at a period of 120 days.

The outcome of the simulation very well reproduces WOCS 5379 in all regards but one. \citet{2019ApJ...885...45G} use an FUV spectrum to measure the WD mass at $0.42^{+0.02}_{-0.02}M_{\odot}$. The WD mass in our model is only 0.33 $M_{\odot}$.

The essential findings from this modeling are the following:

\bigskip

1) Non-conservative mass transfer is necessary to produce the long-period orbit of WOCS 5379. In this regard, we note that \citet{2011Natur.478..356G} found that 14 of the 16 BSSs in NGC 188 are binaries with orbital periods between 120 days and 3030 days. Long-period orbits are typical, and so perhaps is non-conservative mass transfer.

\bigskip

2) All efforts to model WOCS 5379 with conservative MT failed due to rapidly unstable MT. Non-conservative mass transfer is necessary to avoid a common-envelope evolution.

\bigskip

This said, the stability of the MT in the successful model varies during the MT phase. In fact, at its start the MT is briefly unstable, during which time half of the total mass is transferred and half of the system angular momentum is lost. The rapid expansion of the orbit then stabilizes the mass transfer until its completion with the exhaustion of the primary star envelope.

\bigskip

3) The evolution of the progenitor secondary star into a BSS is primarily one of interior re-structuring in response to increasing mass. Upon completion of the MT, the interior structure of the BSS is the same as that of a single star of the same mass at the same location in the HR diagram. While the interiors of two such stars take dramatically different evolutionary paths, in the end the Vogt-Russell Theorem holds.

\bigskip

4) At least in this case, the above finding suggests that the use of single-star models to determine BSS masses may have some validity (but see \citealt{2015ASSL..413...29M}). However, the use of such models for dating BSSs is not valid.

\bigskip

Indeed, the very concept of a BSS age is complex. We note that the \textbf{formation age} of both stars in WOCS 5379 is the age of the cluster, e.g., 6.5 Gyr. However, the MT creating the BSS occurred only 300 Myr ago, which we call the \textbf{transformation age}. Finally, the interior structure of the BSS matches that of a 1.2 $M_{\odot}$ star with an 
\textbf{evolution age} of 2.1 Gyr.

Searches near the initial parameter space produce similar BSS - WD binaries within the observed uncertainty range of the observations. We have not found other domains of parameter space that reproduce WOCS 5379.

However, there are several key physical assumptions in the model that are not well established. First is the location and process of the mass loss from the system. The model presumes that the mass loss is from the vicinity of the accretor. If we assume the mass loss is from the vicinity of the donor, the resulting orbital expansion is smaller (i.e., the average angular momentum loss per mass lost ${\rm d}J/{\rm d}M$ is larger). To produce a final $P_{\rm orb} \sim 120$ days, the initial $P_{\rm orb}$ is near 40 days. Even with the MT being launched at a wider orbit, the resulting BSS - WD binary is very similar to the BF model, with
the final WD mass remaining the same, still significantly smaller than the measured WD mass. At least in this instance, the uncertainties in the angular momentum loss mechanisms do not alter the final product. Even so, better understanding of non-conservative mass transfer, and particularly the related angular momentum evolution, remains essential to complete of understanding these binary evolution pathways.

Second, WOCS 5379 is an eccentric binary, with e=0.24, while the modeling here is done for a circular orbit. For a progenitor orbital period of 12.7 days, it is likely that the initial orbit would be circularized. In this case, the origin of the current orbital eccentricity is not modeled here. Still, it is possible that the progenitor binary may already have had an eccentric orbit. Using an approximation for MT within an eccentric orbit, a model can still match the observed system (except the WD mass). Indeed, the initial parameters are close to the BF model. That said, the MESA approximation for MT in an eccentric orbit is very simple; more developed physics is needed for credible modeling.

Finally, our assumption that the MT efficiency is constant throughout the mass transfer process does not have a physical basis and is most likely not accurate. We consider here whether loosening this constraint might permit an increased final WD mass.

A possible resolution of the WD discrepancy would be to reduce the average MT rate, allowing the donor star more time to build a larger helium core as the system expands to a 120-day period. In single-star evolution of a 1.2 $M_{\odot}$ giant, the core increases from 0.33 $M_{\odot}$ to 0.4 $M_{\odot}$ in approximately 10 Myr, to be compared to the entire time of MT of 46 Myr in the BF model. So the required increase in MT duration is not implausibly large.

In the BF model nearly half of the mass transfer happens during the initial brief unstable phase. As $\dot{M}_1 \propto {\rm exp}(R_1/R_{\rm RL,1})$, one way to lower the initial mass loss rate is for the donor Roche lobe $R_{\rm RL,1}$ to initially expand more rapidly. This can be achieved by lowering the initial MT efficiency. Once the system is stabilized, a return to a higher MT efficiency will be needed to slow the system expansion rate toward $P=120$ days and to produce the BSS.

Exploring pathways without a fixed MT efficiency will be challenging since their is little physical guidance for a variable MT efficiency. Still the experiment is worthwhile to ascertain whether there are any pathways that can resolve the WD discrepancy.

In summary, the MESA stellar evolution code can model the MT evolution of a close main-sequence binary into a blue straggler - white dwarf system with an orbital period of hundreds of days. The nearly complete reproduction of WOCS 5379 supports recent conjectures that most (but not all) BSSs in old open clusters form through MT processes. 

Of course, the success of one case cannot in itself generalize to the wide variety of progenitor MS binaries and product BSS binaries. The ultimate goal must be the detailed production of an entire cluster BSS population. Very likely this will require further development of the essential physics of the stable mass transfer process, with special attention to the processes of stellar rotation (as yet not included), the processes of system mass loss, and orbital eccentricity evolution, along with complementary understanding of common envelope evolution.

\acknowledgements

MS thanks Chenliang Huang and Philipp Podsiadlowski for useful discussions on MT creation of BSSs. We also thank Aaron Dotter and Robert Farmer for help with MESA. Thoughtful comments from the referee improved the paper. Finally we thank Andrew Nine and Melinda Soares-Furtado for careful reading of the manuscript. RDM and MS acknowledge funding support from NSF AST-1714506 and the Wisconsin Alumni Research Foundation. MS and RHDT also acknowledges funding support from NSF ACI-1663696 and AST-1716436. EML is supported by an NSF Astronomy and Astrophysics Postdoctoral Fellowship under award AST-1801937. 

\bibliography{BSS_paper}

\begin{thebibliography}{}
\expandafter\ifx\csname natexlab\endcsname\relax\def\natexlab#1{#1}\fi
\providecommand{\url}[1]{\href{#1}{#1}}
\providecommand{\dodoi}[1]{doi:~\href{http://doi.org/#1}{\nolinkurl{#1}}}
\providecommand{\doeprint}[1]{\href{http://ascl.net/#1}{\nolinkurl{http://ascl.net/#1}}}
\providecommand{\doarXiv}[1]{\href{https://arxiv.org/abs/#1}{\nolinkurl{https://arxiv.org/abs/#1}}}

\bibitem[{{Althaus} {et~al.}(2013){Althaus}, {Miller Bertolami}, \&
  {C{\'o}rsico}}]{2013A&A...557A..19A}
{Althaus}, L.~G., {Miller Bertolami}, M.~M., \& {C{\'o}rsico}, A.~H. 2013,
  \aap, 557, A19, \dodoi{10.1051/0004-6361/201321868}

\bibitem[{{Andronov} {et~al.}(2006){Andronov}, {Pinsonneault}, \&
  {Terndrup}}]{2006ApJ...646.1160A}
{Andronov}, N., {Pinsonneault}, M.~H., \& {Terndrup}, D.~M. 2006, \apj, 646,
  1160, \dodoi{10.1086/505127}

\bibitem[{{Brogaard} {et~al.}(2018){Brogaard}, {Christiansen}, {Grundahl},
  {Miglio}, {Izzard}, {Tauris}, {Sand quist}, {VandenBerg}, {Jessen-Hansen},
  {Arentoft}, {Bruntt}, {Frandsen}, {Orosz}, {Feiden}, {Mathieu}, {Geller},
  {Shetrone}, {Ryde}, {Stello}, {Platais}, \& {Meibom}}]{2018MNRAS.481.5062B}
{Brogaard}, K., {Christiansen}, S.~M., {Grundahl}, F., {et~al.} 2018, \mnras,
  481, 5062, \dodoi{10.1093/mnras/sty2504}

\bibitem[{{Chen} \& {Han}(2004)}]{2004MNRAS.355.1182C}
{Chen}, X., \& {Han}, Z. 2004, \mnras, 355, 1182,
  \dodoi{10.1111/j.1365-2966.2004.08399.x}

\bibitem[{{Chen} \& {Han}(2008)}]{2008MNRAS.387.1416C}
---. 2008, \mnras, 387, 1416, \dodoi{10.1111/j.1365-2966.2008.13334.x}

\bibitem[{{Davis} {et~al.}(2013){Davis}, {Siess}, \&
  {Deschamps}}]{2013A&A...556A...4D}
{Davis}, P.~J., {Siess}, L., \& {Deschamps}, R. 2013, \aap, 556, A4,
  \dodoi{10.1051/0004-6361/201220391}

\bibitem[{{Eggleton}(1983)}]{1983ApJ...268..368E}
{Eggleton}, P.~P. 1983, \apj, 268, 368, \dodoi{10.1086/160960}

\bibitem[{{Fragione} \& {Antonini}(2019)}]{2019MNRAS.488..728F}
{Fragione}, G., \& {Antonini}, F. 2019, \mnras, 488, 728,
  \dodoi{10.1093/mnras/stz1723}

\bibitem[{{Geller} \& {Mathieu}(2011)}]{2011Natur.478..356G}
{Geller}, A.~M., \& {Mathieu}, R.~D. 2011, \nat, 478, 356,
  \dodoi{10.1038/nature10512}

\bibitem[{{Geller} \& {Mathieu}(2012)}]{2012AJ....144...54G}
---. 2012, \aj, 144, 54, \dodoi{10.1088/0004-6256/144/2/54}

\bibitem[{{Geller} {et~al.}(2008){Geller}, {Mathieu}, {Harris}, \&
  {McClure}}]{2008AJ....135.2264G}
{Geller}, A.~M., {Mathieu}, R.~D., {Harris}, H.~C., \& {McClure}, R.~D. 2008,
  \aj, 135, 2264, \dodoi{10.1088/0004-6256/135/6/2264}

\bibitem[{{Geller} {et~al.}(2009){Geller}, {Mathieu}, {Harris}, \&
  {McClure}}]{2009AJ....137.3743G}
---. 2009, \aj, 137, 3743, \dodoi{10.1088/0004-6256/137/4/3743}

\bibitem[{{Gosnell} {et~al.}(2019){Gosnell}, {Leiner}, {Mathieu}, {Geller},
  {Knigge}, {Sills}, \& {Leigh}}]{2019ApJ...885...45G}
{Gosnell}, N.~M., {Leiner}, E.~M., {Mathieu}, R.~D., {et~al.} 2019, \apj, 885,
  45, \dodoi{10.3847/1538-4357/ab4273}

\bibitem[{{Gosnell} {et~al.}(2014){Gosnell}, {Mathieu}, {Geller}, {Sills},
  {Leigh}, \& {Knigge}}]{2014ApJ...783L...8G}
{Gosnell}, N.~M., {Mathieu}, R.~D., {Geller}, A.~M., {et~al.} 2014, \apjl, 783,
  L8, \dodoi{10.1088/2041-8205/783/1/L8}

\bibitem[{{Gosnell} {et~al.}(2015){Gosnell}, {Mathieu}, {Geller}, {Sills},
  {Leigh}, \& {Knigge}}]{2015ApJ...814..163G}
---. 2015, \apj, 814, 163, \dodoi{10.1088/0004-637X/814/2/163}

\bibitem[{{Hamers} \& {Dosopoulou}(2019)}]{2019ApJ...872..119H}
{Hamers}, A.~S., \& {Dosopoulou}, F. 2019, \apj, 872, 119,
  \dodoi{10.3847/1538-4357/ab001d}

\bibitem[{{Hills} \& {Day}(1976)}]{1976ApL....17...87H}
{Hills}, J.~G., \& {Day}, C.~A. 1976, \aplett, 17, 87

\bibitem[{{Hills} {et~al.}(2015){Hills}, {von Hippel}, {Courteau}, \&
  {Geller}}]{2015AJ....149...94H}
{Hills}, S., {von Hippel}, T., {Courteau}, S., \& {Geller}, A.~M. 2015, \aj,
  149, 94, \dodoi{10.1088/0004-6256/149/3/94}

\bibitem[{{Hobbs} {et~al.}(1990){Hobbs}, {Thorburn}, \&
  {Rodriguez-Bell}}]{1990AJ....100..710H}
{Hobbs}, L.~M., {Thorburn}, J.~A., \& {Rodriguez-Bell}, T. 1990, \aj, 100, 710,
  \dodoi{10.1086/115553}

\bibitem[{{Istrate} {et~al.}(2016){Istrate}, {Marchant}, {Tauris}, {Langer},
  {Stancliffe}, \& {Grassitelli}}]{2016A&A...595A..35I}
{Istrate}, A.~G., {Marchant}, P., {Tauris}, T.~M., {et~al.} 2016, \aap, 595,
  A35, \dodoi{10.1051/0004-6361/201628874}

\bibitem[{{Jacobson} {et~al.}(2011){Jacobson}, {Pilachowski}, \&
  {Friel}}]{2011AJ....142...59J}
{Jacobson}, H.~R., {Pilachowski}, C.~A., \& {Friel}, E.~D. 2011, \aj, 142, 59,
  \dodoi{10.1088/0004-6256/142/2/59}

\bibitem[{{Kafka} \& {Honeycutt}(2003)}]{2003AJ....126..276K}
{Kafka}, S., \& {Honeycutt}, R.~K. 2003, \aj, 126, 276, \dodoi{10.1086/375649}

\bibitem[{{Knigge} {et~al.}(2009){Knigge}, {Leigh}, \&
  {Sills}}]{2009Natur.457..288K}
{Knigge}, C., {Leigh}, N., \& {Sills}, A. 2009, \nat, 457, 288,
  \dodoi{10.1038/nature07635}

\bibitem[{{Leigh} \& {Sills}(2011)}]{2011MNRAS.410.2370L}
{Leigh}, N., \& {Sills}, A. 2011, \mnras, 410, 2370,
  \dodoi{10.1111/j.1365-2966.2010.17609.x}

\bibitem[{{Leiner} {et~al.}(2018){Leiner}, {Mathieu}, {Gosnell}, \&
  {Sills}}]{2018ApJ...869L..29L}
{Leiner}, E., {Mathieu}, R.~D., {Gosnell}, N.~M., \& {Sills}, A. 2018, \apjl,
  869, L29, \dodoi{10.3847/2041-8213/aaf4ed}

\bibitem[{{Leiner} {et~al.}(2019){Leiner}, {Mathieu}, {Vanderburg}, {Gosnell},
  \& {Smith}}]{2019ApJ...881...47L}
{Leiner}, E., {Mathieu}, R.~D., {Vanderburg}, A., {Gosnell}, N.~M., \& {Smith},
  J.~C. 2019, \apj, 881, 47, \dodoi{10.3847/1538-4357/ab2bf8}

\bibitem[{{Leonard}(1996)}]{1996ApJ...470..521L}
{Leonard}, P. J.~T. 1996, \apj, 470, 521, \dodoi{10.1086/177884}

\bibitem[{{Mathieu} \& {Geller}(2009)}]{2009Natur.462.1032M}
{Mathieu}, R.~D., \& {Geller}, A.~M. 2009, \nat, 462, 1032,
  \dodoi{10.1038/nature08568}

\bibitem[{{Mathieu} \& {Geller}(2015)}]{2015ASSL..413...29M}
---. 2015, Astrophysics and Space Science Library, Vol. 413, {The Blue
  Stragglers of the Old Open Cluster NGC 188}, ed. H.~M.~J. {Boffin},
  G.~{Carraro}, \& G.~{Beccari}, 29, \dodoi{10.1007/978-3-662-44434-4_3}

\bibitem[{{McCrea}(1964)}]{1964MNRAS.128..147M}
{McCrea}, W.~H. 1964, \mnras, 128, 147, \dodoi{10.1093/mnras/128.2.147}

\bibitem[{{Meibom} {et~al.}(2009){Meibom}, {Grundahl}, {Clausen}, {Mathieu},
  {Frandsen}, {Pigulski}, {Narwid}, {Steslicki}, \&
  {Lefever}}]{2009AJ....137.5086M}
{Meibom}, S., {Grundahl}, F., {Clausen}, J.~V., {et~al.} 2009, \aj, 137, 5086,
  \dodoi{10.1088/0004-6256/137/6/5086}

\bibitem[{{Paczy{\'n}ski}(1971)}]{1971ARA&A...9..183P}
{Paczy{\'n}ski}, B. 1971, \araa, 9, 183,
  \dodoi{10.1146/annurev.aa.09.090171.001151}

\bibitem[{{Paxton} {et~al.}(2011){Paxton}, {Bildsten}, {Dotter}, {Herwig},
  {Lesaffre}, \& {Timmes}}]{2011ApJS..192....3P}
{Paxton}, B., {Bildsten}, L., {Dotter}, A., {et~al.} 2011, \apjs, 192, 3,
  \dodoi{10.1088/0067-0049/192/1/3}

\bibitem[{{Paxton} {et~al.}(2013){Paxton}, {Cantiello}, {Arras}, {Bildsten},
  {Brown}, {Dotter}, {Mankovich}, {Montgomery}, {Stello}, {Timmes}, \&
  {Townsend}}]{2013ApJS..208....4P}
{Paxton}, B., {Cantiello}, M., {Arras}, P., {et~al.} 2013, \apjs, 208, 4,
  \dodoi{10.1088/0067-0049/208/1/4}

\bibitem[{{Paxton} {et~al.}(2015){Paxton}, {Marchant}, {Schwab}, {Bauer},
  {Bildsten}, {Cantiello}, {Dessart}, {Farmer}, {Hu}, {Langer}, {Townsend},
  {Townsley}, \& {Timmes}}]{2015ApJS..220...15P}
{Paxton}, B., {Marchant}, P., {Schwab}, J., {et~al.} 2015, \apjs, 220, 15,
  \dodoi{10.1088/0067-0049/220/1/15}

\bibitem[{{Paxton} {et~al.}(2018){Paxton}, {Schwab}, {Bauer}, {Bildsten},
  {Blinnikov}, {Duffell}, {Farmer}, {Goldberg}, {Marchant}, {Sorokina},
  {Thoul}, {Townsend}, \& {Timmes}}]{2018ApJS..234...34P}
{Paxton}, B., {Schwab}, J., {Bauer}, E.~B., {et~al.} 2018, \apjs, 234, 34,
  \dodoi{10.3847/1538-4365/aaa5a8}

\bibitem[{{Paxton} {et~al.}(2019){Paxton}, {Smolec}, {Schwab}, {Gautschy},
  {Bildsten}, {Cantiello}, {Dotter}, {Farmer}, {Goldberg}, {Jermyn}, {Kanbur},
  {Marchant}, {Thoul}, {Townsend}, {Wolf}, {Zhang}, \&
  {Timmes}}]{2019ApJS..243...10P}
{Paxton}, B., {Smolec}, R., {Schwab}, J., {et~al.} 2019, \apjs, 243, 10,
  \dodoi{10.3847/1538-4365/ab2241}

\bibitem[{{Perets} \& {Fabrycky}(2009)}]{2009ApJ...697.1048P}
{Perets}, H.~B., \& {Fabrycky}, D.~C. 2009, \apj, 697, 1048,
  \dodoi{10.1088/0004-637X/697/2/1048}

\bibitem[{{Sarajedini} {et~al.}(1999){Sarajedini}, {von Hippel},
  {Kozhurina-Platais}, \& {Demarque}}]{1999AJ....118.2894S}
{Sarajedini}, A., {von Hippel}, T., {Kozhurina-Platais}, V., \& {Demarque}, P.
  1999, \aj, 118, 2894, \dodoi{10.1086/301149}

\bibitem[{{Sepinsky} {et~al.}(2009){Sepinsky}, {Willems}, {Kalogera}, \&
  {Rasio}}]{2009ApJ...702.1387S}
{Sepinsky}, J.~F., {Willems}, B., {Kalogera}, V., \& {Rasio}, F.~A. 2009, \apj,
  702, 1387, \dodoi{10.1088/0004-637X/702/2/1387}

\bibitem[{{Sills} {et~al.}(2005){Sills}, {Adams}, \&
  {Davies}}]{2005MNRAS.358..716S}
{Sills}, A., {Adams}, T., \& {Davies}, M.~B. 2005, \mnras, 358, 716,
  \dodoi{10.1111/j.1365-2966.2005.08809.x}

\bibitem[{{Soberman} {et~al.}(1997){Soberman}, {Phinney}, \& {van den
  Heuvel}}]{1997A&A...327..620S}
{Soberman}, G.~E., {Phinney}, E.~S., \& {van den Heuvel}, E.~P.~J. 1997, \aap,
  327, 620.
\newblock \doarXiv{astro-ph/9703016}

\bibitem[{{Tauris} \& {van den Heuvel}(2006)}]{2006csxs.book..623T}
{Tauris}, T.~M., \& {van den Heuvel}, E.~P.~J. 2006, {Formation and evolution
  of compact stellar X-ray sources}, Vol.~39, 623--665

\bibitem[{{Tian} {et~al.}(2006){Tian}, {Deng}, {Han}, \&
  {Zhang}}]{2006A&A...455..247T}
{Tian}, B., {Deng}, L., {Han}, Z., \& {Zhang}, X.~B. 2006, \aap, 455, 247,
  \dodoi{10.1051/0004-6361:20064879}

\bibitem[{{Woods} {et~al.}(2012){Woods}, {Ivanova}, {van der Sluys}, \&
  {Chaichenets}}]{2012ApJ...744...12W}
{Woods}, T.~E., {Ivanova}, N., {van der Sluys}, M.~V., \& {Chaichenets}, S.
  2012, \apj, 744, 12, \dodoi{10.1088/0004-637X/744/1/12}

\end{thebibliography}
\end{CJK*}
\end{document}